\documentclass[12pt, a4paper]{article}
\usepackage[height=25cm,width=16cm]{geometry}
\usepackage{feynmp}
\usepackage{amsmath}
\usepackage{ascmac}
\usepackage{dcolumn}
\usepackage{bm,here}
\usepackage{subfig}
\usepackage{comment}
\usepackage{ifpdf}
\usepackage{slashed}
\usepackage{colortbl}
\usepackage{color}
\usepackage{comment}
\usepackage{braket}
\usepackage[mathscr]{eucal}
\usepackage[sort&compress, numbers, merge]{natbib}

\usepackage{cancel}
\usepackage{subfiles}
\ifpdf
\usepackage{graphicx}     
\usepackage[bookmarksopen,colorlinks=true,linkcolor=bblue,citecolor=ppink,urlcolor=ppink]{hyperref}
\else     
\usepackage[dvipdfmx]{graphicx}     
\usepackage{tikz-feynhand}　
\usepackage[dvipdfmx,bookmarksopen,colorlinks=true,linkcolor=bblue,citecolor=ppink,urlcolor=ppink]{hyperref}
\fi

\usepackage{multicol}
\definecolor{red}{rgb}{1,0,0}
\definecolor{ppink}{rgb}{0.921545,0.440586,0.687243}
\definecolor{bblue}{rgb}{0.400000,0.400000,1.000000}
\usepackage[charter]{mathdesign}
\usepackage{soul}
\usepackage{wrapfig}

\newcommand{\Slash}[1]{{\ooalign{\hfil/\hfil\crcr$#1$}}}

\newcommand\blfootnote[1]{%
	\begingroup
	\renewcommand\thefootnote{}\footnote{#1}%
	\addtocounter{footnote}{-1}%
	\endgroup
}
\begin{document}

\begin{titlepage}

\begin{flushright}
    \hfill IPMU21-xxx \\
\end{flushright}

\begin{center}

	\vskip 1.5cm
	{\Large \bf Decay of the Mediator Particle at Threshold}

	\vskip 2.0cm
	{\large Shigeki Matsumoto$^*$\blfootnote{$^*$shigeki.matsumoto@ipmu.jp},
			Yu Watanabe$^\dagger$\blfootnote{$^\dagger$yu.watanabe@ipmu.jp}, \\ [.3em]
			Yuki Watanabe$^\ddagger$\blfootnote{$^\ddagger$yuki.watanabe@ipmu.jp}
			and
			Graham White$\$$\blfootnote{$^{\$}$graham.white@ipmu.jp} }
			
	\vskip 2.0cm
	{\sl Kavli IPMU (WPI), UTIAS,
    University of Tokyo, Kashiwa, Chiba 277-8583, Japan} \\ [.3em]
	
    \vskip 3.5cm
    \begin{abstract}
        \noindent
        A light mediator particle is often predicted in the dark sector scenario, which weakly interacts with the standard model (SM) particles. The weakness of the interaction is usually described by a small coupling; however, the small coupling does not always guarantee the weakness of the interaction. When the mass of the mediator particle lies in a threshold region, the so-called threshold singularity may emerge, and then the perturbative calculation fails. This singularity causes several effects, e.g., the mixing between the mediator particle and bound states, the Sommerfeld effect on the final state of the mediator particle decay, etc. Taking the minimal model of the vector mediator particle decaying mainly into the SM particles as an example, we develop a method to describe the singularity quantitatively. We also calculate some physical quantities using this method, such as the lifetime of the mediator particle, and find that those could be significantly altered compared with the result of the perturbative calculation.
    \end{abstract}
			
\end{center}
		
\end{titlepage}

\tableofcontents
\newpage
\setcounter{page}{1}

\section{Introduction}
\label{sec: intro}

With the discovery of the Higgs boson in 2012 at the LHC experiment\,\cite{ATLAS:2012yve, CMS:2012qbp}, the standard model (SM) of particle physics had completed not only from a theoretical but also from an experimental perspective. It has brilliantly explained the high-precision experimental results which have been carried out since then and has established a strong position as the most reliable theory we are at hand describing our world. However, while the SM describes many phenomena around us, it is far from the theory of everything. Many things that cannot be explained within this framework are known. One such problem, for example, is related to the free parameters of the SM\,\cite{Li:2011sd, Draper:2016pys, Croon:2019kpe, Peccei:2006as}: dark energy, electroweak scale, gauge couplings, strong CP, etc. There are 20 parameters in the SM, and they cannot be determined from the theory itself and must be measured from various experiments. No one knows how to predict their values from the first principle. Another is about the initial conditions of the universe\,\cite{Martin:2018ycu, Trodden:1998ym}: flatness and isotropic of the universe, baryon asymmetry of the universe, etc. How was the initial condition for the equations giving the evolution of the universe determined? The other is concerned with the experimental anomalies\,\cite{King:2003jb, Bertone:2004pz, Jegerlehner:2009ry}: neutrino masses and mixings, dark matter, muon $g-2$, etc. The SM does not explain these experimental results, and some extension of the SM is required. Building a model beyond the SM is necessary to solve the above puzzles, and it is one of the most important aspects of modern particle physics.

Attempts to find physics beyond the SM (BSM) have already been started before the construction of the LHC. In that era, the mainstream idea was that the origin of solving the problems lies in a high-energy region that we have not yet reached. The best example of this is supersymmetry\,\cite{Martin:1997ns}. Supersymmetry predicts a new particle called superpartner for each SM particle, and its mass, from consistency with various experimental results, was expected to be around $\mathcal{O}(1)$\,TeV. Therefore, it was thought that if supersymmetry exists, it would be sufficiently testable in the LHC experiment. However, the experimental data so far is not favorable for the existence of such a new particle\,\cite{Adam:2021rrw, ATLAS:2022ihe}. Under this circumstance, the so-called hidden sector scenario is attracting attention. In this scenario, we assume that, in addition to the SM sector, an unknown sector exists that contains the mechanism solving (some of) the above problems not in a high-energy region but, like the SM, around the electroweak scale or below. The two sectors must weakly contact each other to avoid severer constraints obtained by various collider experiments. Such a weak contact is guaranteed, e.g., by separating the two sectors by a large energy barrier; the two sectors are only connected through higher dimensional operators originating in some very heavy particles or rare tunneling effects induced by the instanton effect. Such a scenario is called the hidden valley scenario\,\cite{Strassler:2006im}. However, to go from the valley we live in, the SM sector, to the other valley, the hidden sector, we must climb this very high-energy mountain. Hence, accessing the other sector without high-energy experiments is usually challenging.

In contrast, as a more phenomenological and bottom-up approach, the dark sector scenario with a light mediator is recently attracting more attention\,\cite{Essig:2013lka, Alexander:2016aln}. In this scenario, the SM and dark sectors are connected by a light particle called the mediator. The mediator particle is usually assumed to be weakly interacting with the SM particles. Such a dark sector scenario is motivated by various reasons. For instance, if we consider a thermal dark matter scenario assuming that the dark matter is a fermion in the dark sector, a light mediator particle is naturally introduced to make the theory renormalizable and to explain the dark matter density observed today via the freeze-out mechanism\,\cite{Matsumoto:2018acr, Compagnin:2022elr}. Another example is that the light mediator particle itself plays the role of dark matter\,\cite{Graham:2015rva}. Moreover, a light mediator particle is sometimes required to explain experimental anomalies such as the muon $g-2$\,\cite{Chen:2015vqy, Mohlabeng:2019vrz}. In any case, unlike the hidden valley scenario, these light mediator particles are not difficult to produce in high-luminosity collider experiments and to be produced in the early universe. Hence, these particles are now important targets in ongoing and near-future collider experiments\,\cite{Beacham:2019nyx, Alekhin:2015byh}, and cosmological observations\,\cite{Bernal:2017kxu, Fradette:2014sza}.

The lightness of the mass and, in particular, the weakness of the interaction of the mediator particle gives a unique implication for collider physics and cosmology: A weakly interacting particle could be long-lived, giving a distinct signature at collider experiments\,\cite{Alimena:2019zri, Lee:2018pag}. On the other hand, such a particle could be a light relic in the early universe, affecting the cosmological evolution of the universe via its late-time decay or as a light new degree of freedom changing the universe's expansion rate\,\cite{Berger:2016vxi, Dolgov:2002wy}. The weakness of the interaction is usually described by a small coupling. However, the small coupling does not always guarantee the weakness of the interaction. When the mass of the mediator particle lies in the threshold region, the so-called threshold singularity emerges\,\cite{Hisano:2004ds}. Then, the perturbative calculation, which was implicitly assumed in the above implication, fails. This singularity causes several effects, e.g., the mixing between the mediator particle and various bound states, the Sommerfeld effect on the final state of the mediator decay, etc., and could drastically change the implication. In this paper, taking the minimal vector mediator model as an example, assuming it mainly decays into SM particles, we propose a method to describe the singularity quantitatively. Moreover, we evaluate how much influence it has on physical quantities, such as the total decay width of the mediator particle, and find that the prediction could be altered significantly compared with that of the perturbative calculation.

This paper is organized as follows. We first introduce the minimal model of the vector mediator particle in section\,\ref{sec: setup}, and calculate the decay width of the mediator particle perturbatively. Next, in section\,\ref{sec: decay widths}, taking the decay channels into a pair of muons and a pair of bottom quarks as examples, we consider the decay of the mediator particle at their threshold regions and, using the so-called potential non-relativistic method\,\cite{Pineda:1997bj, Brambilla:1999xf}, describe threshold singularities emerged. Here, we see that the Sommerfeld effect on the final state of the decay is systematically incorporated utilizing the method, while also seeing the "sign” of the mixing between the mediator particle and bound states composed of final state particles. So, in the following section (section\,\ref{sec: mixings}), we further develop the method to describe the effect of the mixing quantitatively. Then, we calculate a few physical quantities, e.g., the total decay width of the mediator particle. Finally, in section\,\ref{sec: summary}, we summarize our discussion.

\section{Mediator particle}
\label{sec: setup}

The mediator particle plays a role in connecting dark sector particles with those of the standard model (SM). Among various possibilities, as an example, we consider a vector mediator particle and discuss the threshold singularity on its decay. Below, we introduce the minimal model of the vector mediator that describes its interactions with the SM particles.

\subsection{The minimal model}

We consider the minimal model of a real vector mediator that is singlet under the SM gauge symmetry. It describes interactions between the mediator particle and the SM particles:
\begin{eqnarray}
    {\cal L}_V = {\cal L}_{\rm SM} - \frac{1}{4} (V_{\mu \nu})^2 + \frac{1}{2} M_V^2 (V_\mu)^2 - \frac{\xi}{2} V_{\mu \nu}\,B^{\mu \nu} + \frac{\lambda_{VH}}{2} (V_\mu)^2 |H|^2 + \frac{\lambda_V}{4!} (V_\mu)^4 + \cdots + {\cal L}_{\rm DS},
    \label{eq: lagrangian V}
\end{eqnarray}
where ${\cal L}_{\rm SM}$ is the SM lagrangian, while ${\cal L}_{\rm DS}$ is the dark sector lagrangian involving interactions between the mediator particle and dark sector particles. We do not specify details of ${\cal L}_{\rm DS}$, as it does not play an essential role in the following discussion.\footnote{We assume that the mediator particle decays visibly, i.e., mainly into SM particles, throughout this paper.} The field describing the vector mediator is denoted by $V_\mu$ with its field strength tensor to be $V_{\mu \nu} \equiv \partial_\mu V_\nu - \partial_\nu V_\mu$, while $B_{\mu \nu}$ is that of the hyper-charge gauge boson $B_\mu$, and $H$ is the SM Higgs doublet. The vector field $V_\mu$ was initially introduced as a gauge field associated with a new U(1) symmetry imposed on the theory. This symmetry is assumed to be broken by an Abelian Higgs field $\varphi$, making the vector field $V_\mu$ massive. The above lagrangian is obtained by integrating the physical mode of the Abelian Higgs field, assuming it is heavier than other particles.

The kinetic mixing term proportional to $\xi$ in ${\cal L}_V$ makes the vector mediator $V_\mu$ mixed with the SM gauge fields $B_\mu$ and also $W^3_\mu$ via the electroweak symmetry breaking, where $W^3_\mu$ is the electrically neutral component of the SU(2)$_L$ gauge field multiplet. After taking the unitary gauge, the quadratic terms of the fields $W^3_\mu$, $B_\mu$, and $V_\mu$ are obtained as follows:
\begin{align}
    &{\cal L}_V \supset
    \frac{1}{2} (W^3_\mu, B_\mu, V_\mu)
    \left[ (\Box g^{\mu \nu} - \partial^\mu \partial^\nu)\,{\cal K} + {\cal M}\,g^{\mu \nu} \right]
    (W^3_\nu, B_\nu, V_\nu)^T,
    \nonumber \\
    & {\cal K} \equiv
    \begin{pmatrix} 1 & 0 & 0 \\ 0 & 1 & \xi \\ 0 & \xi & 1 \end{pmatrix},
    \qquad
    {\cal M} \equiv
    \begin{pmatrix} g^2 v_H^2/4 & -g g' v_H^2/4 & 0 \\ -g g' v_H^2/4 &  g^{\prime 2} v_H^2/4 & 0 \\ 0 & 0 & M_V^2 + \lambda_{VH} v_H^2/2 \end{pmatrix},
\end{align}
where $g$ and $g'$ are the couplings of the SU(2)$_L$ and U(1)$_Y$ gauge interactions of the SM, respectively, while $v_H \simeq$ 246\,GeV is the vacuum expectation value of the Higgs field $H$. The redefinition of the vector mediator and the SM gauge fields, as well as the diagonalization of the above mass matrix ${\cal M}$, give mass eigenstates having canonical kinetic terms,
\begin{align}
    &
    X {\cal K} X^T = {\bf 1},
    \qquad
    X {\cal M} X^T = {\rm diag}(m_Z^2, 0, m_{A'}^2),
    \qquad
    (Z_\mu, A_\mu, A'_\mu)^T = (X^{-1})^T (W^3_\mu, B_\mu, V_\mu)^T,
    \nonumber \\
    & \qquad\qquad
    X \simeq
    \begin{pmatrix}
        \cos \theta_W & -\sin \theta_W & \displaystyle \xi \frac{\sin \theta_W m_Z^2}{m_Z^2 - m_{A'}^2} \\
        \sin \theta_W & \cos \theta_W & 0 \\
        \displaystyle -\xi \frac{\cos \theta_W \sin \theta_W m_Z^2}{m_Z^2 - m_{A'}^2} & \displaystyle \xi \frac{m_{A'}^2 - \cos^2 \theta_W m_Z^2}{m_Z^2 - m_{A'}^2} & 1
    \end{pmatrix},
\end{align}
assuming the parameter $\xi \ll 1$ with $m_{A'}$ being the mass of the vector mediator particle $A'$.

Using the mass eigenstates of the particles $Z_\mu$, $A_\mu$ and $A'_\mu$, i.e., $Z$ boson, photon, and the vector mediator particle, respectively, the above lagrangian in eq.\,(\ref{eq: lagrangian V}) is written as
\small
\begin{align}
    &{\cal L}_V = \frac{1}{2} A'_\mu \left[ (\Box + m_{A'}^2)\,g^{\mu \nu} - \partial^\mu \partial^\nu \right] A'_\nu
    +\frac{\epsilon\,e\,m_Z^2}{m_Z^2 - m_{A'}^2} \left[ A'_\mu J^\mu_{\rm EM} + F_{A'\,W^\dagger\,W} \right]
    -\frac{\epsilon\,(g'/c_W)\,m_{A'}^2}{m_Z^2 - m_{A'}^2} A'_\mu J^\mu_Y
    + \cdots,
    \nonumber \\
    &J^\mu_{\rm EM} =
    \sum_{i =1}^3 Q^{(\ell)}_i \bar{\ell}_i \gamma^\mu \ell_i
    +\sum_{i = 1}^6 Q^{(q)}_i \bar{q}_i \gamma^\mu q_i,
    \label{eq: interactions V}
    \nonumber \\
    &F_{A'\,W^\dagger\,W} = i
    \left[
        (\partial^\mu W^{\nu \dagger} - \partial^\nu W^{\mu \dagger}) W_\mu A'_\nu
        -(\partial^\mu W^\nu - \partial^\nu W^\mu) W_\mu^\dagger A'_\nu
        +(\partial^\mu A^{\prime \nu} - \partial^\nu A^{\prime \mu}) W_\mu^\dagger W_\nu
    \right],
    \nonumber \\
    &J^\mu_Y = \sum_{i = 1}^3
    \left[
        Y^{(\ell)}_{L_i} \bar{\ell}_i \gamma^\mu P_L \ell_i
        +Y^{(\ell)}_{R_i} \bar{\ell}_i \gamma^\mu P_R \ell_i 
        +Y^{(\nu)}_i \bar{\nu}_i \gamma^\mu \nu_i
    \right]
    +\sum_{i = 1}^6
    \left[
        Y^{(q)}_{L_i} \bar{q}_i \gamma^\mu P_L q_i
        +Y^{(q)}_{R_i} \bar{q}_i  \gamma^\mu P_R q_i
    \right],
\end{align}
\normalsize
where $\epsilon \equiv \xi \cos\theta_W$, $c_W = \cos \theta_W$, and $e$ is the electromagnetic coupling. On the other hand, $Q^{(\ell)}_i$ and $Q^{(q)}_i$ are the electric charges of the charged lepton $\ell_i$ and the quark $q_i$, while $Y^{(\ell)}_{L_i}$, $Y^{(\ell)}_{R_i}$, $Y^{(\nu)}_i,$ $Y^{(q)}_{L_i}$ and $Y^{(q)}_{R_i}$ are the hyper-charges of the left-handed charged lepton $P_L \ell_i$, the right-handed charged lepton $P_R \ell_i$, the neutrino $\nu_i$, the left-handed quark $P_L q_i$ and the right-handed quark $P_R q_i$, respectively. Here, ``$\dots$'' represents other interactions of $A'$ with the $Z$ and Higgs bosons (i.e., $A' Z h$, $(A')^2 h$, $A' Z h^2$, etc.) in addition to those in ${\cal L}_{\rm SM}$ and ${\cal L}_{\rm DS}$.

\subsection{Decay of the mediator particle}
\label{subsec: perturbative decays}

We briefly summarize the decay channels and corresponding partial decay widths of the mediator particle $A'$ within a perturbative calculation. As seen in the lagrangian\,(\ref{eq: lagrangian V}), the vector mediator particle interacts with the SM particles through the mixing with the hyper-charge gauge boson, so it decays mainly into a pair of the SM particles via the electromagnetic and the weak neutral currents. In particular, when the mediator is lighter enough than the electroweak scale, i.e., the $Z$ boson mass, it dominantly decays via the electromagnetic current, for the decay via the neutral current is suppressed by a factor $m_{A'}^2/m_Z^2$, as seen in eq.\,(\ref{eq: interactions V}).

When the mass of the mediator particle is less than twice the electron mass $2 m_e$, the mediator particle decays mainly into three photons via loop diagrams that an electron is propagating. The corresponding partial decay width is $\Gamma(A' \to 3\gamma) \simeq 4.7 \times 10^{-8} \epsilon^2 \alpha^4 m_{A'}^9/m_e^8$\,\cite{Pospelov:2008jk}, with $\alpha$ being the fine structure constant. The mediator particle can also decay into a pair of neutrinos via the neutral weak current interaction, though its partial decay width is suppressed by a factor of $m_{A'}^4/m_Z^4 \lesssim 10^{-20}$.\footnote{This decay channel dominates when the vector mediator model is constructed based on, e.g., U(1)$_{B-L}$.} On the other hand, when the mediator particle is heavier than twice the electron mass, it can decay into a pair of electrons, namely, an electron and a positron. The partial decay width of the mediator particle into a pair of the SM leptons, i.e., electron, muon, or tau lepton pairs, is generally obtained as follows:
\begin{align}
    \Gamma_0(A' \to \ell_i^- \ell_i^+)
    \simeq \epsilon^2 \frac{ \alpha m_{A'} }{ 3 }
    \left( 1 + 2 \frac{ m_{\ell_i}^2 }{ m_{A'}^2 } \right)
    \sqrt{ 1 - \frac{ 4m_{\ell_i}^2 }{ m_{A'}^2 } }.
    \label{eq: perturbative decay into a lepton pair}
\end{align}

When the mediator particle is heavier than twice the pion mass $2m_\pi$, it can also decay hadronically. A convenient way to evaluate the total hadronic decay width of the mediator particle is the use of the so-called $R$-ratio obtained by the result of collider experiments\,\cite{ParticleDataGroup:2020ssz}, i.e., $R = \sigma(e^- e^+ \to {\rm all~hadrons}; m_{A'})/\sigma(e^- e^+ \to \mu^- \mu^+; m_{A'})$ with $\sigma(e^- e^+ \to \cdots; m_{A'})$ being the annihilation cross-section between an electron and a positron with the center of mass energy of $m_{A'}$. Then, the hadronic decay width is given by the  following formula\,\cite{Fabbrichesi:2020wbt}:
\begin{align}
    \Gamma_0(A' \to {\rm all~hadrons})
    \simeq \epsilon^2 \frac{ \alpha m_{A'} }{ 3 }
    \left( 1 + 2 \frac{ m_\mu^2 }{ m_{A'}^2 } \right)
    \sqrt{ 1 - \frac{ 4m_\mu^2 }{ m_{A'}^2 } }\,R,
    \label{eq: perturbative decay into hadrons}
\end{align}
assuming that $m_{A'}$ is enough smaller than the electroweak scale. On the other hand, the partial decay width of each hadronic channel is robustly computed using the chiral lagrangian method when $m_{A'} \leq$ 500\,MeV\,\cite{Coogan:2021sjs}. In contrast, when $m_{A'} \geq$ a few GeV, the width can be evaluated as the sum of those into various pairs of quarks based on a perturbative calculation\,\cite{Buschmann:2015awa}. The partial decay width of the mediator particle into a pair of SM quarks is obtained as $\Gamma_0(A' \to q_i \bar{q}_i) \simeq 3 Q_i^{(q)} \Gamma_0(A' \to \ell_i^- \ell_i^+) |_{m_{\ell_i} \to m_{q_i}}$ with $m_{q_i}$ being the quarks-mass in the final state. When the mass of the mediator particle is intermediate, i.e., 500\,MeV $\leq m_{A'} \leq$ a few GeV, the partial decay width is evaluated based on the exclusive cross-section measurement of the $e^- e^+$ annihilation process into a corresponding hadronic final state\,\cite{Buschmann:2015awa}, though extracting data from the measurement often gives large systematic uncertainties.

\section{Decay of the mediator particle at threshold}
\label{sec: decay widths}

In this section, we consider the decay of the vector mediator particle at threshold regions, i.e., the regions where the sum of final state (daughter) particles in mass is close to the mass of the vector mediator (parent) particle. As examples, we consider the decay of the mediator particle at the threshold region of a muon pair and the region of a bottom quark pair. The method developed here can also be applied to other threshold regions.

\subsection{Decay into a pair of muons at the threshold}
\label{subsec: decay into leptons}

Let us first consider the decay of the mediator particle at the threshold region of a muon pair, i.e., $m_{A'} \simeq 2 m_\mu$ with $m_\mu \simeq$ 106\,MeV being the mass of the muon. The electromagnetic interaction between the muons in this mass region causes a long-range force, leading to the so-called threshold singularity. We can consider its effect on the decay width using the so-called potential non-relativistic (NR) lagrangian method\,\cite{Pineda:1997bj, Brambilla:1999xf}: Integrating out all the fields except those of the mediator particle $A'(x)$ and the non-relativistic part of the muon $\mu_{\rm NR}(x)$\footnote{The non-relativistic part of the muon field is defined as $\mu_{\rm NR}(x) \equiv \int_{[{\rm NR}]} d^4p/(2\pi^4)\,\tilde{\mu}(p)\,e^{-i p x}$, where $\tilde{\mu}(p)$ is the Fourier coefficient of the muon field $\mu(x)$, while $[{\rm NR}]$ is the domain of the integration that corresponds to the almost on-shell and non-relativistic muon, i.e. $p^0 = \pm m_\mu + \delta p^0$ with $\delta p^0$ being ${\cal O}({\bf p}^2/m_\mu) \ll m_\mu$.} from the lagrangian\,(\ref{eq: interactions V}), and expanding $\mu_{\rm NR}(x)$ in terms of its velocity as
\begin{align}
    \mu_{\rm NR}(x) =
    \begin{pmatrix}
        e^{- i m_\mu x^0} \eta_\mu(x) + i e^{i m_\mu x^0} [\vec{\nabla} \cdot \vec{\sigma}\,\xi_\mu(x)]/(2 m_\mu) + \cdots \\
        e^{i m_\mu x^0} \xi_\mu(x) - i e^{-i m_\mu x^0} [\vec{\nabla} \cdot \vec{\sigma}\,\eta_\mu(x)]/(2 m_\mu) + \cdots
    \end{pmatrix},
    \label{eq: NR expansion}
\end{align}
with $\sigma$ being the Pauli matrix, gives us the non-relativistic lagrangian as follows:
\small
\begin{align}
    \mathcal{L}^{({\rm NR})}_V
    &\simeq
    \frac{1}{2} A'_\mu \left[ (\Box + m_{A'}^2)\,g^{\mu\nu} - \partial^\mu \partial^\nu \right] A'_\nu
    + \eta_\mu^\dagger \left( i\partial_{x^0} + \frac{\nabla^2}{2m_\mu} + i\frac{\Gamma_\mu}{2} \right) \eta_\mu
    + \xi_\mu^\dagger \left( i\partial_{x^0} - \frac{\nabla^2}{2m_\mu} - i\frac{\Gamma_\mu}{2} \right) \xi_\mu
    \nonumber \\
    &+
    \frac{1}{2} \int d^4y \frac{\alpha \delta(x^0 - y^0)}{|\vec{x}-\vec{y}|}
    \left[ \eta_\mu^\dagger(x)\,\vec{\sigma}\,\xi_\mu(y) \right] \cdot \left[ \xi_\mu^\dagger(y)\,\vec{\sigma}\,\eta_\mu(x) \right]
    \nonumber \\
    &+
    i\frac{\pi \alpha^2}{3 m_\mu^2} \left[ \eta_\mu^\dagger\,\vec{\sigma}\,\xi_\mu \right] \cdot \left[ \xi_\mu^\dagger\,\vec{\sigma}\,\eta_\mu \right]
    + \epsilon e\,\vec{A}' \cdot [e^{2 i m_\mu x^0} \eta_\mu^\dagger\,\vec{\sigma}\,\xi_\mu + e^{-2 i m_\mu x^0}\xi_\mu^\dagger \vec{\sigma} \eta_\mu].
    \label{eq: NR lagrangian}
\end{align}
\normalsize
$\eta_\mu\,(\eta_\mu^\dagger)$ is the operator annihilating\,(creating) a non-relativistic muon $\mu^-$, while $\xi_\mu\,(\xi_\mu^\dagger)$ creates\,(annihilates) its anti-particle $\mu^+$. In the above lagrangian, we drop several terms describing other states not directly coupled to the mediator field, such as $\eta_\mu^\dagger(x) \xi_\mu(y)$. Moreover, $\Gamma_\mu$ is the decay width of the muon, which is obtained by integrating out the $W$ boson and lighter SM fermions from the original lagrangian in ${\cal L}_V$. It contributes to the decay width of the $\mu^- \mu^+$ two-body state. On the other hand, the other imaginary term (the first term in the third line) in the above lagrangian describes the annihilation between the muons $\mu^-$ and $\mu^+$, and it also contributes to the decay width of the $\mu^- \mu^+$ two-body state.

After introducing the two-body fields describing the $\mu^- \mu^+$ two-body system that directly couples to the vector mediator field $A'$, and integrating the component fields $\eta_\mu$ and $\xi_\mu$ in the above lagrangian\,(\ref{eq: NR lagrangian}), we have the potential NR lagrangian for the two-body system:
\small
\begin{align}
    \mathcal{L}^{({\rm pNR})}_V
    &\simeq 
    -\frac{1}{2} \vec{A}'\,(\Box +m_{A'}^2)\,\vec{A}'
    + \int d^3r\,\vec{\Phi}^\dagger(\vec{r},x) \left(i\partial_{x^0} + \frac{\nabla_x^2}{4m_\mu} + \frac{\nabla_r^2}{m_\mu} - V(\vec{r}) \right)  \vec{\Phi}(\vec{r},x) \nonumber \\
    &+
    \sqrt{2} \epsilon e \vec{A}'\,\left[ e^{2i m_\mu x^0} \vec{\Phi}^\dagger (\vec{0}, x) + e^{-2i m_\mu x^0} \vec{\Phi}(\vec{0},x) \right],
    \label{eq: pNR lagrangian}
\end{align}
\normalsize
where $\vec{A}'$ is the spatial components of the vector mediator field, while $\vec{\Phi}(\vec{r},x)$ and $\vec{\Phi}^\dagger(\vec{r},x)$ are the fields (operators) annihilating and creating the $\mu^- \mu^+$ two-body system with the total spin of one (the spin-triplet state), respectively. Here, $\vec{r}$ and $x$ are the relative and center-of-mass coordinates of the two-body system. The potential $V(\vec{r})$ is given by
\begin{align}
    V(\vec{r}) = -\frac{\alpha}{r} -i\Gamma_\mu - i\frac{2\pi \alpha^2}{3m_\mu^2}\delta(\vec{r}),
\end{align}
where the real part of the potential represents the Coulomb force acting between $\mu^-$ and $\mu^+$, while the imaginary (absorption) part describes the so-called "decay width" of the $\mu^- \mu^+$ two-body system, which originates in the decay of the muons and the annihilation between them into a pair of electrons (an electron and a positron), as mentioned above.

Now we calculate the decay width of the mediator particle at the threshold region of a muon pair. To make the calculation intuitive, we utilize the expansion of the two-body field $\Phi(\vec{r},x)$ in terms of the solutions of the Schr\"{o}dinger equation, i.e., the equation of motion of the two-body field, that describes the relative motion between the muons. Treating the imaginary part of the potential perturbatively, we can expand the field as follows:
\begin{align}
    \vec{\Phi}(\vec{r},x) = \sum_{n,\,\ell,\,m}\,\vec{B}_{n \ell m}(x)\,\psi^{(B)}_{n \ell m}(\vec{r})
    + \int \frac{dk}{2\pi}\,\sum_{\ell,\,m}\,\vec{C}_{k \ell m}(x)\,\psi^{(C)}_{k \ell m}(\vec{r}),
    \label{eq: expansion}
\end{align}
where $\psi^{(B)}_{n \ell m}(\vec{r})$ is the solution describing a bound state with $n$, $\ell$, and $m$ being the principal, azimuthal, and magnetic quantum numbers, respectively, and the coefficient $\vec{B}_{n \ell m}(x)$ is the field (operator) annihilating the corresponding bound state. On the other hand, $\psi^{(C)}_{k \ell m}(\vec{r})$ is the solution describing a continuum state with the wave number $k = (m_\mu E)^{1/2}$, and the quantum numbers $\ell$, $m$, while the coefficient $\vec{C}_{k \ell m}(x)$ is the field annihilating the continuum state. The solutions $\psi^{(B)}_{n \ell m}(\vec{r})$ and $\psi^{(C)}_{k \ell m}(\vec{r})$ are analytically obtained thanks to the Coulomb potential\,\cite{Landau_Lifshitz}, and the substitution of the expansion for the lagrangian\,(\ref{eq: pNR lagrangian}) gives
\small
\begin{align}
    \mathcal{L}^{(\text{pNR})}_V
    &\simeq
    -\frac{1}{2} \vec{A}'\,(\Box + m_{A'}^2)\,\vec{A}'
    +\sum_n \vec{B}^\dagger_{n00} \left[ i\partial_0 + \frac{\nabla^2}{4m_\mu} + \frac{\alpha^2 m_\mu}{4n^2} + i\Gamma_\mu + i\frac{\alpha^5 m_\mu}{12n^3} \right] \vec{B}_{n00}
    \nonumber \\
    &+
    \int \frac{dk}{2\pi}\,\vec{C}^\dagger_{k00} \left[ i\partial_0 + \frac{\nabla^2}{4m_\mu} - \frac{k^2}{m_\mu} \right] \vec{C}_{k00}
    +\sqrt{2} \epsilon e \vec{A}'
    \left[
        e^{2 i m_\mu x^0} \sum_n \frac{1}{\sqrt{\pi}} \left(\frac{\alpha m_\mu}{2n}\right)^{3/2} \vec{B}^\dagger_{n00}
    \right.
    \nonumber \\
    &\qquad\qquad\qquad\qquad\qquad\qquad
    +\left.
        e^{2 i m_\mu x^0} \int \frac{dk}{2\pi} \left\{ \frac{\alpha m_\mu k}{1 - \exp\,(-\pi \alpha m_\mu/k)} \right\}^{1/2}
        \vec{C}^\dagger_{k00}
    + \text{h.c.}
    \right],
    \label{eq: pNR lagrangian 2}
\end{align}
\normalsize
where we omitted writing the fields with the quantum number $\ell \neq 0$: because their wave functions, $\psi^{(B)}_{n \ell m}(\vec{r})$ and $\psi^{(C)}_{k \ell m}(\vec{r})$, vanish at the origin, i.e., at $\vec{r} = 0$, and the fields do not couple to the vector mediator particle directly, as seen in eq.\,(\ref{eq: pNR lagrangian}). Moreover, we also dropped the imaginary parts of the kinetic terms, i.e., the decay widths, for the fields describing the continuum states, for as do not play an essential role in the following discussion.\footnote{On the other hand, the imaginary parts for the fields describing the bound states play a role, as seen in the following discussion. For the fields with a larger principal number $n$, there is an additional contribution to the imaginary part describing the deexcitation to lower bound states by emitting a photon(s). We do not include the contribution in the above lagrangian, as it does not change an essential part of our discussion.}

With the above lagrangian\,(\ref{eq: pNR lagrangian 2}), the "decay width" of the vector mediator particle into a pair of muons is obtained using the self-energy of the mediator particle $\Sigma\,(q)$ as follows:
\begin{align}
    \Gamma\,(A' \to \mu^- \mu^+) =  -\frac{1}{m_{A'}}{\rm Im}\,\left[ \Sigma\,(m_{A'}) \right],
\end{align}
where $\Sigma\,(q)$ is defined as $\int d^4x\,e^{i q x} \braket{0| T[A^i(x) A^j(0)] |0} \equiv i \delta^{ij}/[q^2 - m_{A'}^2 - \Sigma\,(q)]$. Both the bound and the continuum states contribute to $\Sigma\,(q)$, and the decay width is obtained as
\small
\begin{align}
    \Gamma\,(A' \to \mu^- \mu^+)
    = \sum_n \frac{\epsilon^2 \pi \alpha^4 m_\mu^3}{n^3 m_{A'}} f(m_{A'}; M^{(\mu\mu)}_n, \Gamma^{(\mu\mu)}_n/2) 
    +\frac{2 \pi \epsilon^2 \alpha^2 m_\mu^2 \theta(m_{A'} - 2m_\mu)}{m_{A'}\,\left[1 - \exp\,(-\pi\alpha/\sqrt{m_{A'}/m_\mu - 2})\right]},
    \label{eq: decay width into leptons}
\end{align}
\normalsize
where the first term on the right-hand side is the contribution from the bound states, and the function $f(x; x_0, \gamma) \equiv (1/\pi) \gamma/[(x - x_0)^2 + \gamma^2]$ is the Breit-Wigner function with $M^{(\mu\mu)}_n \equiv 2m_\mu - \alpha^2 m_\mu/(4 n^2)$ and $\Gamma^{(\mu\mu)}_n \equiv 2\Gamma_\mu + \alpha^5 m_\mu/(6 n^3)$ being the mass and the width of the $n$-th excited bound state composed of $\mu^- \mu^+$ pair. On the other hand, the second term is the contribution from the continuum states of the pair. When the effect of the long-range force is weakened, i.e., $\pi\alpha/\sqrt{m_{A'}/m_\mu - 2} \to 0$, the second term gives $(\epsilon^2 \alpha m_{A'}/\sqrt{2}) \sqrt{1 - 2m_\mu/m_{A'}}$, and it coincides with the partial decay width shown in eq.\,(\ref{eq: perturbative decay into a lepton pair}) with taking $m_{A'} \to 2m_\mu$.

\begin{figure}[t]
    \centering
    \includegraphics[keepaspectratio, scale=0.35]{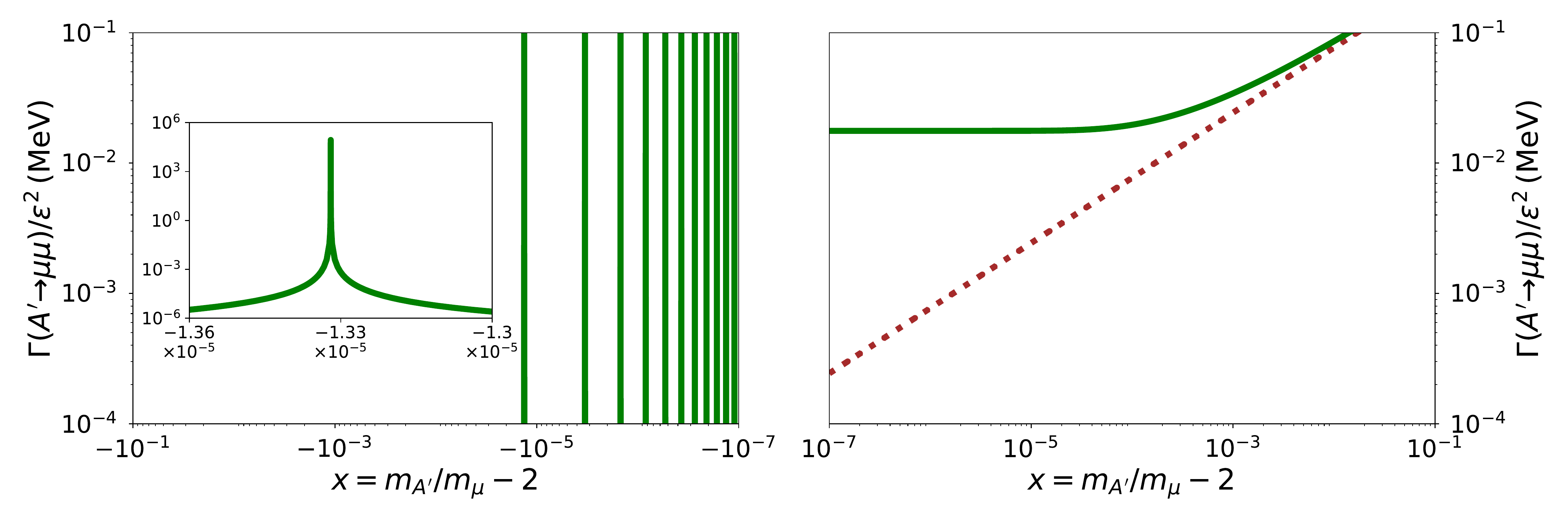}
    \caption{\small \sl A solid green line shows the partial decay width of the mediator particle into a pair of muons in the range of $x \le 0$ (left panel) and the range of $x \ge 0$ (right panel). The partial decay width obtained by the perturbative calculation given in eq.\,(\ref{eq: perturbative decay into a lepton pair}) is also shown in the right panel by a dotted brown line for comparison purposes. See the text and appendix\,\ref{app: matching} for more details.}
    \label{fig: decay width into leptons}
\end{figure}

The decay width in eq.\,(\ref{eq: decay width into leptons}) is shown by a solid green line in Fig.\,\ref{fig: decay width into leptons} as a function of $x \equiv m_{A'}/m_\mu - 2$.\footnote{The decay width plotted in the $x \ge 0$ region is not precisely the same as in eq.\,(\ref{eq: decay width into leptons}). The width involves higher partial wave contributions in addition to the s-wave contribution in eq.\,(\ref{eq: decay width into leptons}). See appendix\,\ref{app: matching} for more details. However, the difference between them is negligibly small in the range of $x$ adopted in the figure.} The width obtained by the perturbative calculation in eq.\,(\ref{eq: perturbative decay into a lepton pair}) is also shown by a dotted brown line for comparison purposes. As seen in the right panel of the figure, the effect of the long-range force becomes sizable when $x \leq 10^{-2}$. It is indeed${\cal O}(10\,\%)$ level at $x \sim 10^{-2}$ and is more than ${\cal O}(100\,\%)$ level when $x \gtrsim 10^{-4}$. On the other hand, as seen in the left panel of the figure, several Breit-Wigner resonances can be found in the region of $x \leq 0$. Since their widths are very narrow, their peak values become enormous. However, it does not mean that the decay width becomes very large when the mass of the mediator particle is in the vicinity of the resonances. It instead indicates a sign of the mixing between the vector mediator particle and (one of) the bound states composed of $\mu^- \mu^+$. In the next section (section\,\ref{sec: mixings}), we will discuss how the mixing is quantitatively taken into account.

\subsection{Decay into a pair of bottom quarks
at the threshold}
\label{subsec: decay into quarks}

Next, let us consider the decay of the mediator particle at the threshold region of a bottom quark pair, i.e., $m_{A'} \simeq 2 m_b \simeq 10$\,GeV with $m_b$ being the bottom quark mass. In this case, not only the electromagnetic interaction but also the strong interaction causes a long-range force acting between the bottom quarks. Hence, it leads to a stronger threshold singularity on the partial decay width of the mediator particle. As in the case discussed in the previous subsection (i.e., $A' \to \mu^- \mu^+$), the effect of the long-range forces on the partial decay width can be considered using the method of the potential non-relativistic lagrangian\,\cite{Pineda:1997bj, Brambilla:1999xf}. Then, the effective lagrangian, in this case, is eventually obtained as follows:
\small
\begin{align}
    \mathcal{L}^{({\rm pNR})}_V
    &\simeq 
    -\frac{1}{2} \vec{A}'\,(\Box +m_{A'}^2)\,\vec{A}'
    + \int d^3r\,\vec{\chi}^\dagger(\vec{r},x) \left(i\partial_{x^0} + \frac{\nabla_x^2}{4m_b} + \frac{\nabla_r^2}{m_b} - V(\vec{r}) \right)  \vec{\chi}(\vec{r},x) \nonumber \\
    &+
    \sqrt{2/3}\,\epsilon e \vec{A}'\,\left[ e^{2i m_b x^0} \vec{\chi}^\dagger (\vec{0}, x) + e^{-2i m_b x^0} \vec{\chi}(\vec{0},x) \right],
    \label{eq: pNR lagrangian QCD LO}
\end{align}
\normalsize
where $\vec{\chi}(\vec{r},x)$ and $\vec{\chi}^\dagger(\vec{r},x)$ are the fields (operators) annihilating and creating the two-body state of a bottom quark pair with the total spin of one (the spin-triplet state). The potential inside the kinetic term of the fields is given by $V(\vec{r}) = -4\alpha_s/(3r) - \alpha/(9r) - i \Gamma_b - 20 i (\pi^2 - 9) \alpha_s^3 \delta(\vec{r})/(81 m_b^2)$ at leading order, with $g_s = (4\pi\alpha_s)^{1/2}$ being the strong coupling constant. Here, the first and the second terms of the potential describe the long-range forces from the strong and the electromagnetic interactions, respectively. The third term is the imaginary (absorption) part originating in the decay of the bottom quarks, while the fourth term is the one originating in the annihilation between the bottom quarks into three gluons.

On the other hand, the kinetic term of the two-body fields in the above lagrangian, i.e., the Schr\"{o}dinger equation of the two-body states, is substantially corrected by the strong interaction. The correction is more substantial when higher excited states are concerned, i.e., when $m_{A'}$ closer to the threshold is considered because their wave functions are more spread to a larger space and become more sensitive to the infrared QCD behavior. One of the typical consequences of the correction is the color-confinement; no continuum $b \bar{b}$ states appear due to the correction, unlike the $\mu^- \mu^+$ system in section\,\ref{subsec: decay into leptons}. One of the methods to take the correction into account is the use of the Schr\"{o}dinger equation with the potential that is improved theoretically by higher-order QCD corrections with the help of experimental data phenomenologically\,\cite{Segovia:2016xqb, Chaturvedi:2018tjr, Kher:2022gbz}. Another method is only the use of experimental data. Because the quantum numbers of the two-body states coupling to the vector mediator particle are the same as those of the final states in $e^- e^+$ collision, the $R$-ratio and spectroscopic data of various mesons can be directly used to evaluate the wave functions of the two-body states. We will therefore adopt the latter method, as explained below in more detail.

Assuming that the potential, including the QCD correction, is approximated to be spherically symmetric, the two-body fields can be expanded in the same manner as eq.\,(\ref{eq: expansion}), 
\begin{align}
    \vec{\chi}(\vec{r},x) = \sum_{n,\,\ell,\,m}\,\vec{D}_{n \ell m}(x)\,\psi^{(D)}_{n \ell m}(\vec{r}) = \sum_{n,\,\ell,\,m}\,\vec{D}_{n \ell m}(x)\,R_{n \ell m}(r) Y_{\ell m}(\theta, \phi),
\end{align}
with $r$, $\theta$, and $\phi$ being the spherical coordinates. The radial part of the wave function is denoted by $R_{n \ell m}(r)$, while $Y_{\ell m}(\theta, \phi)$ is the spherical harmonic function. Substituting this expansion into the effective lagrangian with the potential involving the correction gives
\small
\begin{align}
    \mathcal{L}^{({\rm pNR})}_V
    &\simeq
    -\frac{1}{2}\vec{A}'\,(\Box + m_{A'}^2)\,\vec{A}'
    + \sum_n \vec{D}^\dagger_{n00} \left[i \partial_0 + \frac{\nabla^2}{4m_b} - E^{(b b)}_n + \frac{i}{2} \Gamma^{(b b)}_n \right] \vec{D}_{n00}
    \nonumber \\
    &+ \sqrt{2/3} \epsilon e \vec{A}'
    \left[
        e^{2 i m_b x^0} \sum_n \frac{R_{n00}(0)}{\sqrt{4\pi}} \vec{D}_{n00}^\dagger
        +
        e^{- 2 i m_b x^0} \frac{R_{n00}(0)}{\sqrt{4\pi}} \vec{D}_{n00}
    \right],
    \label{eq: pNR lagrangian QCD}
\end{align}
\normalsize
where $\vec{D}_{n00}(x)$ and $\vec{D}_{n00}^\dagger(x)$ are the fields (operators) annihilating and creating the ($n - 1$)-th excited $b \bar{b}$ bound state with the quantum number $J^{PC} = 1^{-\,-}$ (the total spin of one and the orbital angular momentum of zero), i.e., the $\Upsilon(n S)$ meson, respectively, with $M^{(b b)}_n \equiv 2m_b + E^{(b b)}_n$ and $\Gamma^{(b b)}_n$ being the mass and the total decay width of the bound state. Hence, those are directly obtained by the spectroscopic data of the meson\,\cite{ParticleDataGroup:2020ssz}. On the other hand, the wave-function of the ($n - 1$)-th excited state at the origin, i.e., $R_{n00}(0)$, is obtained by various branching fractions of the meson decay. Using the wave function at the origin, the partial decay widths of the upsilon meson decaying into a pair of electrons, a gluon pair associated with a photon, and three gluons are obtained by the formulae\,\cite{Segovia:2016xqb, PhysRevD.37.3210, Brambilla:2001xy, Brambilla:2002nu, Vairo:2003gh}:
\begin{align}
    \Gamma\,[ \Upsilon(n S) \to e^- e^+ ]
    &= \frac{4\,\alpha^2 |R_{n00}(0)|^2}{9\,[M^{(b b)}_n]^2} \left( 1 - \frac{16}{3\pi} \alpha_s \right),
    \label{eq: branching fraction 1}
    \\
    \Gamma\,[ \Upsilon(n S) \to \gamma g g ]
    &= \frac{8\,(\pi^2 - 9)\,\alpha \, \alpha_s^2 |R_{n 00}(0)|^2}{81 \pi m_b^2}
    \left( 1 - \frac{7.4 }{\pi} \alpha_s \right),
    \label{eq: branching fraction 2}
    \\
    \Gamma\,[ \Upsilon(n S)\to 3g]
    &= \frac{10\,(\pi^2 - 9)\,\alpha_s^3 |R_{n 00}(0)|^2}{81 \pi m_b^2}
    \left( 1 - \frac{4.9}{\pi} \alpha_s \right),
    \label{eq: branching fraction 3}
\end{align}
where QCD corrections at the 1-loop level are included. The value of the wave-function $R_{n00}(0)$ is obtained by comparing the branching fractions of the meson decay, Br\,$[\Upsilon(n S) \to e^- e^+ ] \equiv \Gamma\,[ \Upsilon(n S) \to e^- e^+ ]/\Gamma^{(b b)}_n$, etc., with experimental data. The values of $M^{(b b)}_n$, $\Gamma^{(b b)}_n$, and $|R_{n00}(0)|^2$ for up to the third excited ($n = 4$) bound state are obtained as follows:
\begin{table}[h!]
    \centering
    \begin{tabular}{c|ccc}
        & $M^{(b b)}_n$\,(GeV) & $\Gamma^{(b b)}_n$\,(MeV) & $|R_{n00}(0)|^2$\,(GeV$^3$) \rule[-2mm]{0mm}{1mm}
        \\
        \hline
        $\Upsilon(1S)$  & $\simeq 9.46$ & $\simeq 0.054$ & $\simeq 8.7$ \\
        $\Upsilon (2S)$ & $\simeq 10.0$ & $\simeq 0.032$ & $\simeq 4.1$. \\
        $\Upsilon (3S)$ & $\simeq 10.4$ & $\simeq 0.020$ & $\simeq 2.1$ \\
        $\Upsilon (4S)$ & $\simeq 10.6$ & $\simeq 21$\,\,\,\,\,~ & $\simeq 2.1 $ \\
        \hline
    \end{tabular}
\end{table}

\begin{figure}[t]
    \centering
 \includegraphics[keepaspectratio, scale=0.35]{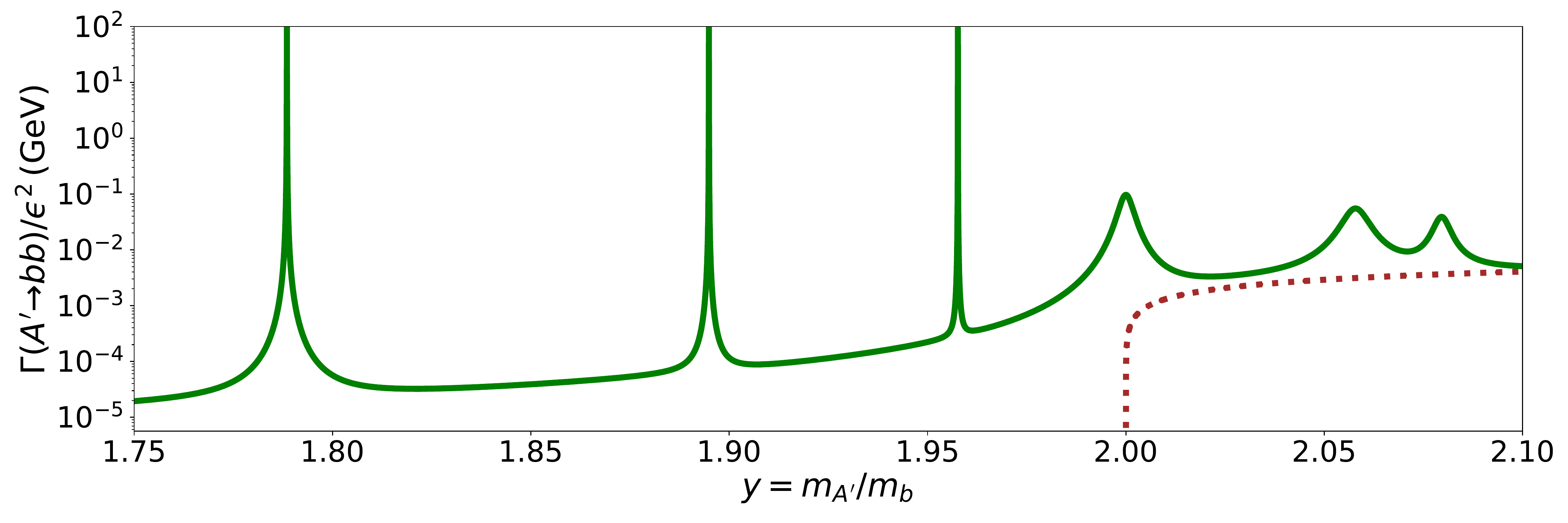}
    \caption{\small \sl A solid green line shows the partial decay width of the mediator particle into a pair of bottom quarks. The decay width obtained by a naive quark-parton model calculation is also shown by a dotted brown line for comparison purposes. See the text and appendix\,\ref{app: matching 2} for more details.}
    \label{fig: decay width into quarks}
\end{figure}

Using the same method discussed in section\,\ref{subsec: decay into leptons}, the effective lagrangian in eq.\,(\ref{eq: pNR lagrangian QCD}) gives the "partial decay width" of the mediator particle into a pair of bottom quarks as follows:
\begin{align}
    \Gamma\,( A' \to b \bar{b} )
    = \sum_n \frac{2 \epsilon^2 \pi \alpha}{3 m_{A'}}|R_{n00}(0)|^2f(m_{A'};M^{(b b)}_n,\Gamma^{(b b)}_n/2),
    \label{eq: decay width into hadrons}
\end{align}
where $f(x; x_0, \gamma)$ is the Breit-Wigner function in eq.\,(\ref{eq: decay width into leptons}). A solid green line shows this decay width in Fig.\,\ref{fig: decay width into quarks} as a function of $y \equiv m_{A'}/m_b$ with $m_b$ being $5.3$\,GeV.\footnote{The value of the bottom quark mass, $m_b \simeq 5.3$\,GeV, is obtained by the comparison between the prediction in eqs.\,(\ref{eq: branching fraction 1})--(\ref{eq: branching fraction 3}) and experimental data. Moreover, the decay width plotted in the figure is not precisely the same as the one in eq.\,(\ref{eq: decay width into hadrons}). It involves an additional contribution from continuum $B \bar{B}$ states at $m_{A'} > M^{(b b)}_4$, which is obtained by the R-ratio (the 3-loop QCD prediction\,\cite{Chetyrkin:2000zk}). See appendix\,\ref{app: matching 2} for more details.} The decay width obtained by a naive quark-parton model calculation is also shown by a dotted brown line for comparison purposes. Several Breit-Wigner resonances corresponding to the upsilon mesons can be found in the solid green line, which is very different from the result (the dotted brown line) of the naive quark-parton model calculation due to the $b\bar{b}$ bound states. The decay width becomes enormous when the mass of the mediator particle is in the vicinity of the resonances. However, as addressed in section\,\ref{subsec: decay into leptons}, it does not mean that the width behaves as shown in the figure, but it indicates the existence of the mixing between the mediator particle and the bound states. We will discuss it in detail in the next section.

\section{Mixing between mediator particle and bound states}
\label{sec: mixings}

As mentioned in the previous section, we consider the mixing effect between the vector mediator particle and bound states (discrete states) composed of a muon pair or a bottom quark pair. This section mainly discusses how it affects the total decay width of the mediator particle. On the other hand, since the mixing also affects the mediator particle production at colliders, we also address it using a $e^- e^+$ collider (e.g., Belle\,II, ILC, FCC-ee, CEPC, etc.) as an example. The mixing of $A'$ with $\mu^- \mu^+$ bound states (muonic atoms) is discussed in section\,\ref{subsec: mixing with muons}, while that with $b \bar{b}$ bound states (upsilon mesons) is in section\,\ref{subsec: mixing with bottoms}.

\subsection{Mixing with muonic atoms}
\label{subsec: mixing with muons}

Using the same method of the potential NR lagrangian developed in section\,\ref{subsec: decay into leptons}, we can obtain the effective lagrangian that describes interactions among the vector mediator particle, the $\mu^- \mu^+$ bound states (the muonic atoms), the electron, and the positron as follows:
\small
\begin{align}
    \mathcal{L}^{(\text{pNR})}_V
    &\simeq
    -\frac{1}{2} \vec{A}'\,(\Box + m_{A'}^2)\,\vec{A}'
    +\bar{\psi}_e\,(i \Slash{D} - m_e)\,\psi_e
    +\sum_n \vec{B}^\dagger_{n00} \left[ i\partial_0 + \frac{\nabla^2}{4m_\mu} - E_n^{(\mu \mu)} \right] \vec{B}_{n00}
    \label{eq: pNR lagrangian with mixing}
    \\
    &+
    \sqrt{2} e
    \left[
        \epsilon \vec{A}'
        +
        i e \int d^4y {\cal G}^{(\gamma)}(x - y) \bar{\psi}_e(y) \vec{\gamma} \psi_e(y)
    \right]
    \cdot
    \left[
        \sum_n \frac{e^{2 i m_\mu x^0}}{\sqrt{\pi}} \left(\frac{\alpha m_\mu}{2n}\right)^{\frac{3}{2}} \vec{B}^\dagger_{n00}
        +
        \text{h.c.}
    \right] + \cdots,
    \nonumber
\end{align}
\normalsize
where $\psi_e$ is the electron field with $\Slash{D} = \Slash{\partial} + i e \vec{A} \cdot \vec{\gamma} - i \epsilon e \vec{A}' \cdot \vec{\gamma}$ and $m_e$ being its covariant derivative and the electron mass, respectively. On the other hand, the binding energy of the $(n - 1)$-th excited bound state is denoted by $E_n^{(\mu \mu)}$, which is $E_n^{(\mu \mu)} = -\alpha^2 m_\mu/(4 n^2)$ at leading order calculation, and ${\cal G}^{(\gamma)}(x) \equiv i \int d^4q\,e^{-i q x}/[(2 \pi)^4\,(q^2 + i0^+)]$ is the photon propagator. Finally, "$\cdots$" involves other terms of the SM particles and the mediator particle.\footnote{Unlike the lagrangian in eq.\,(\ref{eq: pNR lagrangian 2}), the decay widths of the bound states are not included in eq.\,(\ref{eq: pNR lagrangian with mixing}); the widths are obtained by integrating other SM particles such as the electron, the $W$ boson, and neutrinos, etc.} To describe the mixing between the mediator particle and the bound states, it is helpful to rewrite the non-relativistic field $\vec{B}_{n00}(x)$ by the form generally used in the relativistic field theory:
\begin{align}
    \vec{\phi}_n(x) \equiv
    \frac{1}{\sqrt{4m_\mu}}
    \left[
        e^{-2 i m_\mu x^0} \vec{B}_{n00}(x) + e^{2 i m_\mu x^0} \vec{B}^\dagger_{n00}(x)
    \right].
\end{align}
With this vector field, the potential NR lagrangian\,(\ref{eq: pNR lagrangian with mixing}) is given by the following form:
\begin{align}
    \mathcal{L}^{(\text{pNR})}_V
    &\simeq
    -\frac{1}{2} \vec{A}'\,(\Box + m_{A'}^2)\,\vec{A}'
    +\bar{\psi}_e\,(i \Slash{D} - m_e)\,\psi_e
    -\sum_{n} \frac{1}{2} \vec{\phi}_n (\Box + m_{\phi_n}^2) \vec{\phi}_n
    \nonumber \\
    &+ \frac{e\,m_\mu^2}{\sqrt{\pi}}
    \left[
        \epsilon \vec{A}'
        +
        i e \int d^4y {\cal G}^{(\gamma)}(x - y) \bar{\psi}_e(y) \vec{\gamma} \psi_e(y)
    \right]
    \cdot
    \sum_n \left(\frac{\alpha}{n}\right)^{\frac{3}{2}} \vec{\phi}_n + \cdots,
    \label{eq: pNR lagrangian with mixing modified}
\end{align}
where the mass of the bound state field $\vec{\phi}_n(x)$ is denoted by $m_{\phi_n} \equiv 2 m_\mu + E_n^{(\mu \mu)}$. 

When the mass of the vector mediator particle ($m_{A'}$) becomes very close to that of a muonic atom ($m_{\phi_n}$), these two states are expected to be highly mixed with each other. Then, the mass eigenstates are given by the linear combinations of the two states rather than themselves, which are obtained by diagonalizing the mass matrix as follows:
\begin{align}
    \mathcal{L}^{(\text{pNR})}_V \,\supset\,
	-\frac{1}{2}
	\begin{pmatrix} \vec{A}' & \vec{\phi}_n \end{pmatrix}
	\begin{pmatrix} m_{A'}^2 & m_{A' \phi_n}^2 \\ m_{A' \phi_n}^2 & m_{\phi_n}^2 \end{pmatrix}
	\begin{pmatrix} \vec{A}' \\ \vec{\phi}_n \end{pmatrix}
	\,=\,
	-\frac{1}{2}
	\begin{pmatrix} \vec{\tilde{A}}' & \vec{\tilde{\phi}}_n \end{pmatrix}
	\begin{pmatrix} m_{\tilde{A}'}^2 & 0 \\ 0 & m_{\tilde{\phi}_n}^2 \end{pmatrix}
	\begin{pmatrix} \vec{\tilde{A}}' \\ \vec{\tilde{\phi}}_n \end{pmatrix},
\end{align}
where the off-diagonal element is defined as $m_{A' \phi_n}^2 = -\epsilon e m_\mu^2 (\alpha/n)^{3/2}/\sqrt{\pi}$. The mixing matrix diagonalizing the mass matrix and relating the states $(\vec{A}', \vec{\phi}_n)$ and $(\vec{\tilde{A}}', \vec{\tilde{\phi}}_n)$ is given by
\begin{align}
	\begin{pmatrix} \vec{\tilde{A}}' \\ \vec{\tilde{\phi}}_n \end{pmatrix} =
	\begin{pmatrix} \cos\Theta_n & -\sin\Theta_n \\ \sin\Theta_n & \cos\Theta_n \end{pmatrix}
	\begin{pmatrix} \vec{A}' \\ \vec{\phi}_n \end{pmatrix}.
\end{align}
Here, the mass eigenstates and the mixing angle are obtained as $m_{\tilde{A}'}^2, m_{\tilde{\phi}_n}^2 = \{(m_{A'}^2 + m_{\phi_n}^2) \pm (m_{A'}^2 - m_{\phi_n}^2)
[1 + 4m_{A' \phi_n}^4/(m_{A'}^2 - m_{\phi_n}^2)^2]^{1/2}\}/2$, and $\tan(2\Theta_n) = 2 m_{A' \phi_n}^2/(m_{\phi_n}^2 - m_{A'}^2)$ with the domain of $-\pi/4 \leq \Theta_n \leq \pi/4$. In such a case, the effective lagrangian\,(\ref{eq: pNR lagrangian with mixing modified}) is written as
\small
\begin{align}
    \mathcal{L}^{(\text{pNR})}_V
    &\simeq
    -\frac{1}{2} \vec{\tilde{A}}'\,(\Box + m_{\tilde{A}'}^2)\,\vec{\tilde{A}}'
    +\bar{\psi}_e\,(i \Slash{\partial} - e\,\vec{A} \cdot \vec{\gamma} - m_e)\,\psi_e
    -\frac{1}{2} \vec{\tilde{\phi}}_n (\Box + m_{\tilde{\phi}_n}^2) \vec{\tilde{\phi}}_n
    \label{eq: pNR lagrangian with mixing modified 2}
    \\
    &+ e \vec{\tilde{A}}' \cdot \int d^4y 
    \left[
        \epsilon \cos \Theta_n\,\delta (x - y)
        - i e \sin \Theta_n \frac{m_\mu^2}{\sqrt{\pi}} \left( \frac{\alpha}{n} \right)^{3/2} {\cal G}^{(\gamma)}(x - y)
    \right] \bar{\psi}_e(y) \vec{\gamma} \psi_e(y)
    \nonumber \\
    &+ e \vec{\tilde{\phi}}_n \cdot \int d^4y 
    \left[
        \epsilon \sin \Theta_n\,\delta (x - y)
        + i e \cos \Theta_n \frac{m_\mu^2}{\sqrt{\pi}} \left( \frac{\alpha}{n} \right)^{3/2} {\cal G}^{(\gamma)}(x - y)
    \right] \bar{\psi}_e(y) \vec{\gamma} \psi_e(y)
    + \cdots,
    \nonumber
\end{align}
\normalsize
where "$\cdots$" describes other interactions, e.g., those of the bound states, $\vec{\phi}_{m \neq n}(x)$.

\subsubsection{Decay width of the vector mediator particle}

Using the lagrangian\,(\ref{eq: pNR lagrangian with mixing modified 2}), we can calculate the decay width of the vector mediator particle for the case that the mediator particle degenerates with one of the muonic atoms $\vec{\phi}_n$ in mass. The partial decay width of the mediator particle into an electron pair is given by
\begin{align}
    \Gamma_n(\tilde{A}' \to e^- e^+)
    = \frac{\alpha m_{A'}}{3}
    \left[
        \epsilon \cos \Theta_n + \frac{e}{\sqrt{\pi}}\left(\frac{\alpha}{n}\right)^{3/2}\left(\frac{m_\mu}{m_{A'}}\right)^2 \sin \Theta_n 
    \right]^2,
    \label{eq: decay width with mixing 1}
\end{align}
where we neglect the electron mass in the final state and use the approximation $m_{\tilde{A}'} \simeq m_{A'}$ assuming $\epsilon \ll 1$. It is seen in eq.\,(\ref{eq: decay width with mixing 1}) that, even if the $\epsilon$ is very suppressed, the decay width is enhanced when the mixing angle $\Theta_n$ is large enough. Its maximum value is given by $\Gamma_n^{\rm (max)}(\tilde{A}' \to e^- e^+) = \alpha m_{A'}[\epsilon^2 + 4 \alpha^4 (m_\mu / m_{A'})^2/n^3]/3$. On the other hand, it is also seen that the two contributions in the parenthesis are destructively interfered with at a negative $\Theta_n$, making the partial decay width vanish. When the mixing angle is small, the decay width becomes the same as that of the perturbative calculation given in eq.\,(\ref{eq: perturbative decay into a lepton pair}), as expected.

\begin{figure}[t]
    \centering
    \includegraphics[keepaspectratio, scale=0.35]{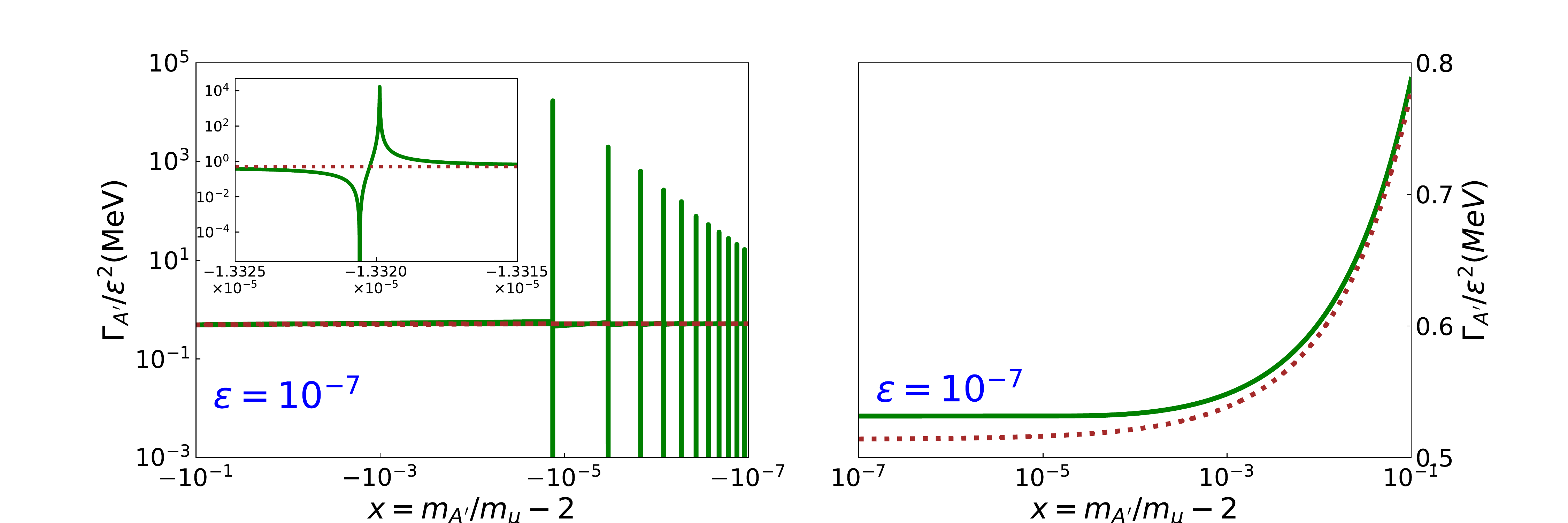}
    \caption{\small \sl A solid green line shows the total decay width of the mediator particle in the range of $x \leq 0$ (left panel) and the range of $x \geq 0$ (right panel), where $x = m_{A'}/m_\mu - 2$. The decay width obtained by the perturbative calculation is also shown in both panels by a dotted brown line.}
    \label{fig: total decay width at mumu}
\end{figure}

The vector mediator particle decays mainly into a pair of electrons in the region of $m_{A'} \leq 2 m_\mu$, while into both pairs of electrons and muons in the region of $m_{A'} \geq 2 m_\mu$.\footnote{When $m_{A'} \simeq m_{\phi_n}$ with $n$ being a large enough, the mediator particle can also decay into lower muonic atoms $\phi_{m < n}$ and photon(s) through the deexcitation of the muonic atom $\phi_n$ into lighter ones. We do not include its contribution in our calculation because it does not change an essential part of our discussion.} Hence, in the vicinity of the threshold region of a muon pair, the total decay width (lifetime) of the mediator particle as a function of the mass $m_{A'}$ is approximately obtained as follows:
\small
\begin{align}
    \Gamma_{A'} \simeq \Gamma_0(A' \to e^- e^+)
    + \sum_n \left[ \Gamma_n(\tilde{A}' \to e^- e^+) - \Gamma_0(A' \to e^- e^+)\right]
    + \Gamma\,(A' \to \mu^- \mu^+)\,\theta(m_{A'} - 2m_\mu),
\end{align}
\normalsize
where $\Gamma_0(A' \to e^- e^+) = \epsilon^2 \alpha m_{A'}/3$ is the partial decay width into an electron pair obtained by the perturbative calculation in eq.\,(\ref{eq: perturbative decay into a lepton pair}), while $\Gamma\,(A' \to \mu^- \mu^+)$ is that given in eqs.\,(\ref{eq: decay width into leptons}) and (\ref{eq: gamma mu mu at app}). The decay width $\Gamma_{A'}$ is depicted in Fig.\,\ref{fig: total decay width at mumu} assuming $\epsilon = 10^{-7}$. It is seen that, when $m_{A'} \leq 2 m_\mu$, the total decay width is enhanced very much in the vicinity of $m_{A'} \simeq m_{\phi_n}$ compared to that of the perturbative calculation, though it remains finite as we take the threshold effect into account through the mixing between $A'$ and $\phi_n$. On the other hand, when $m_{A'} \geq m_{\phi_n}$, the width is slightly enhanced by the threshold effect, as seen in section\,\ref{subsec: decay into leptons}.

The total decay width of the $(n - 1)$-th excited muonic atom $\phi_n$ is also obtained from the lagrangian in eq.\,(\ref{eq: pNR lagrangian with mixing modified 2}). Using the same method as that for $A'$, the width is obtained as
\begin{align}
    \Gamma_{\tilde{\phi}_n} \simeq
    \frac{\alpha m_{\phi_n}}{3} \left[ \frac{e}{\sqrt{\pi}} \left( \frac{\alpha}{n} \right)^{3/2} \left(\frac{m_\mu}{m_{\phi_n}} \right)^2 \cos \Theta_n  - \epsilon \sin \Theta_n \right]^2.
\end{align}
The decay width vanishes at a positive $\Theta_n$. On the other hand, it coincides with the standard prediction of the muonic atom found in Ref.\,\cite{Brodsky:2009gx} when the mixing angle is suppressed.

\subsubsection{Production of the vector mediator particle at \texorpdfstring{$e^- e^+$}{TEXT} colliders}
\label{subsubsec: production at mu mu}

As mentioned at the beginning of this section, the mixing also affects the production of the mediator particle at collider experiments. So, we next consider the direct production of the vector mediator particle at a $e^- e^+$ collider as an example. At given center-of-mass energy of the collision $s^{1/2} \gg m_{A'}$, one of the effective processes to produce the mediator particle is $e^- e^+ \to A' \gamma$ with "$\gamma$" being a photon. When the mediator particle has a mass around twice the muon mass and its decay length is not too long, it is efficiently detected by observing the final state of its decay, i.e., an electron pair. The mass of the mediator particle is then reconstructed by measuring the invariant mass of the pair\,\cite{Graham:2021ggy}. Moreover, when the decay length of the mediator particle is shorter than the size of the detector but longer than ${\cal O}(1)$\,mm, the displaced vertex of the $A'$-decay may be reconstructed from the final state. Then, this enables us to reduce backgrounds and measure the lifetime of $A'$\,\cite{Duerr:2019dmv}.

\begin{figure}[t]
    \centering
    \includegraphics[keepaspectratio, scale=0.35]{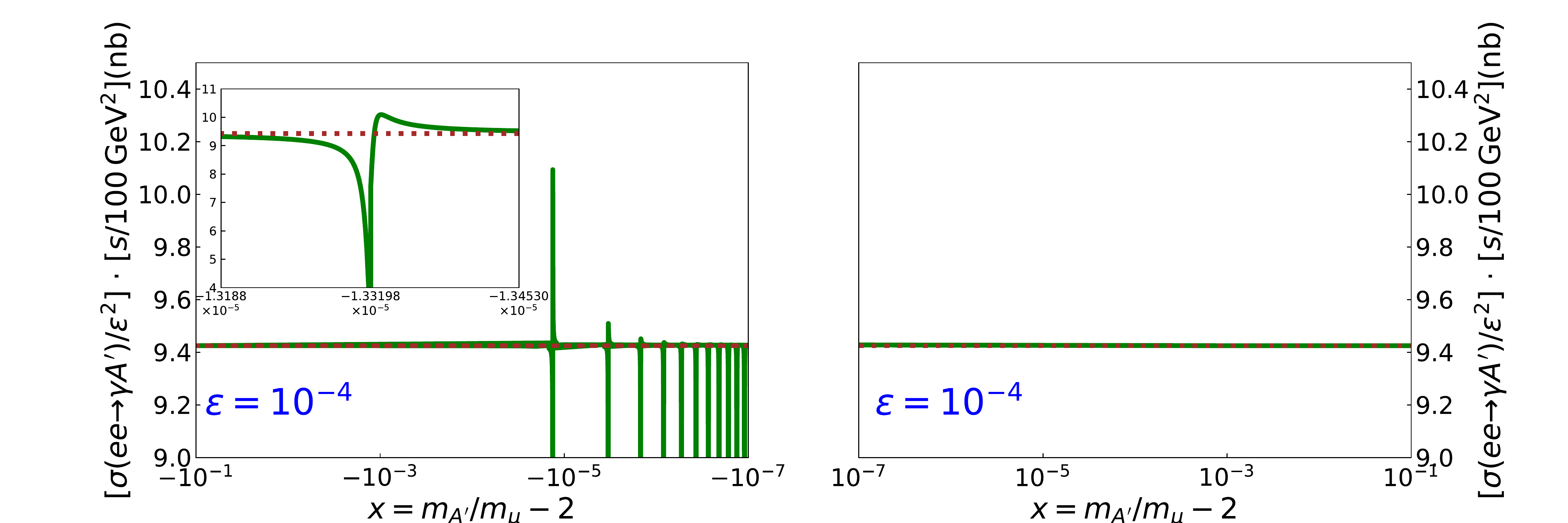}
    \caption{\small \sl The production cross-section of the mediator particle as a function of $x = m_{A'}/m_\mu - 2$, obtained by integrating the differential cross-section\,(\ref{eq: production cross-section mumu}) over the angles $-0.98 \leq \cos\theta \leq 0.98$. The cross-section obtained by the perturbative calculation is also shown by a dotted brown line.}
    \label{fig: production and events at mumu}
\end{figure}

When the vector mediator particle is not highly degenerate with any of the muonic atoms, the differential production cross-section of the mediator particle at the collider is
\begin{align}
    \frac{d \sigma_0 (e^- e^+ \to \gamma A')}{d\cos\theta} \simeq
    \frac{2 \pi \epsilon^2 \alpha^2}{s} \frac{1 + \cos^2\theta}{\sin^2\theta},
    \label{eq: perturbative production cross-section mumu} 
\end{align}
with $\theta$ being the angle between the directions of the mediator particle and the incoming electron. On the other hand, when the mediator particle highly degenerates with one of the muonic atoms, i.e., $m_{A'} \simeq m_{\phi_n}$, the cross-section is obtained using the lagrangian\,(\ref{eq: pNR lagrangian with mixing modified 2}) as
\begin{align}
    \frac{d \sigma_n (e^- e^+ \to \gamma \tilde{A}')}{d\cos\theta} \simeq
    \left[
        \epsilon \cos \Theta_n
        +
        \frac{e}{\sqrt{\pi}} \left( \frac{\alpha}{n} \right)^{3/2} \left(\frac{m_\mu}{m_{A'}} \right)^2 \sin \Theta_n
    \right]^2
    \frac{2 \pi \alpha^2}{s} \frac{1 + \cos^2\theta}{\sin^2\theta}.
\end{align}
When the mixing angle $\Theta_n$ is suppressed enough, the above production cross-section becomes the same as that of the perturbative calculation given in eq.\,(\ref{eq: perturbative production cross-section mumu}). So, the production cross-section of the vector mediator particle as a function of $m_{A'}$ is given as follows:
\begin{align}
    \frac{d \sigma (e^- e^+ \to \gamma A')}{d\cos\theta}
    \simeq
    \frac{d \sigma_0 (e^- e^+ \to \gamma A')}{d\cos\theta}
    +
    \sum_n
    \left[
        \frac{d \sigma_n (e^- e^+ \to \gamma \tilde{A}')}{d\cos\theta}
        -
        \frac{d \sigma_0 (e^- e^+ \to \gamma \tilde{A}')}{d\cos\theta}
    \right].
    \label{eq: production cross-section mumu} 
\end{align}
The production cross-section of the mediator particle, obtained by integrating the above differential cross-section over the scattering angles $-0.98 \leq \cos\theta \leq 0.98$, is depicted in Fig.\,\ref{fig: production and events at mumu} assuming $\epsilon = 10^{-4}$. When $m_{A'} \leq 2 m_\mu$, the mixing slightly enhances the cross-section compared to that of the perturbative calculation at $m_{A'} \simeq m_{\phi_n}$, or suppresses it as the mixing causes a destructive interference between diagrams contributing to the cross-section.

\subsection{Mixing with upsilons}
\label{subsec: mixing with bottoms}

We next consider the mixing effect between the vector mediator particle and bound states composed of a pair of bottom quarks, i.e., upsilon mesons. Using the same method of the potential NR lagrangian discussed in section\,\ref{subsec: decay into quarks}, we obtain the effective lagrangian that describes interactions among the vector mediator particle, the $b \bar{b}$ bound states (upsilon mesons) with $b$ being a bottom quark, and various SM fermions denoted by $f_i$:
\small
\begin{align}
    \mathcal{L}^{({\rm pNR})}_V
    &\simeq
    -\frac{1}{2}\vec{A}'\,(\Box + m_{A'}^2)\,\vec{A}'
    +\sum_i \bar{f}_i\,(i \Slash{D} - m_{f_i})\,f_i
    + \sum_n \vec{D}^\dagger_{n00} \left[i \partial_0 + \frac{\nabla^2}{4m_b} - E^{(b b)}_n \right] \vec{D}_{n00}
    \label{eq: pNR lagrangian with mixing bb}
    \\
    + \sqrt{\frac{2}{3}} & e
    \left[
        \epsilon \vec{A}'
        -
        i e \sum_i \int d^4y\,Q_i\,{\cal G}^{(\gamma)}(x - y) \bar{f}_i(y) \vec{\gamma} f_i(y)
    \right]
    \cdot
    \left[
        \sum_n \frac{e^{2 i m_b x^0} }{\sqrt{4\pi}} R_{n00}(0)\,\vec{D}_{n00}^\dagger
        +
        \text{h.c.}
    \right] + \cdots ,
    \nonumber
\end{align}
\normalsize
where $m_{f_i}$ is the mass of the SM fermion $f_i$. Other interactions, such as those between upsilons and gluons, interactions describing transitions among the upsilons, and other SM interactions, etc., are not explicitly written in the above lagrangian and involved in "$\cdots$" to avoid a verbose description.\footnote{These interactions contribute to, e.g., the total decay widths of the upsilons seen in the lagrangian\,(\ref{eq: pNR lagrangian QCD}), once we integrate out the daughter particles of their decay processes from the lagrangian given above.} As in the case of the previous subsection, to describe the mixing between the mediator particle and the upsilons, it is again helpful to rewrite the non-relativistic field $\vec{D}_{n00}(x)$ in terms of the form used in the relativistic field theory:
\begin{align}
    \vec{\varphi}_n(x) \equiv
    \frac{1}{\sqrt{4m_b}}
    \left[
        e^{-2 i m_b x^0} \vec{D}_{n00}(x) + e^{2 i m_b x^0} \vec{D}^\dagger_{n00}(x)
    \right].
\end{align}
With the vector field $\vec{\varphi}_n(x)$, the potential NR lagrangian\,(\ref{eq: pNR lagrangian with mixing bb}) is written as follows:
\begin{align}
    \mathcal{L}^{(\text{pNR})}_V
    &\simeq
    -\frac{1}{2} \vec{A}'\,(\Box + m_{A'}^2)\,\vec{A}'
    +\sum_i \bar{f}_i\,(i \Slash{D} - m_{f_i})\,f_i
    -\sum_{n} \frac{1}{2} \vec{\varphi}_n (\Box + m_{\varphi_n}^2) \vec{\varphi}_n
    \label{eq: pNR lagrangian with mixing modified bb 1}
    \\
    &+
    e \sqrt{\frac{2m_b}{3\pi}}
    \left[
        \epsilon \vec{A}'
        -
        i e \sum_i \int d^4y\,Q_i\,{\cal G}^{(\gamma)}(x - y) \bar{f}_i(y) \vec{\gamma} f_i(y)
    \right]
    \cdot
    \sum_n R_{n00}(0)\,\vec{\varphi}_n + \cdots,
    \nonumber
\end{align}
where the mass of the bound state field $\vec{\phi}_n(x)$ is denoted by $m_{\varphi_n}^2 \equiv 4m_b(m_b + E_n^{(bb)})$.

As in the case of the mixing between the vector mediator particle $A'$ and muonic atoms discussed in the previous subsection, $A'$ is mixed with one of the upsilon mesons when the mediator mass $m_{A'}$ becomes very close to the upsilon mass $m_{\varphi_n}$. Then, $A'$ and the upsilon states are mixed, and the mass eigenstates are given by their linear combinations as
\begin{align}
    \mathcal{L}^{(\text{pNR})}_V \,\supset \, 
    -\frac{1}{2}
    \begin{pmatrix} \vec{A}' & \vec{\varphi}_n \end{pmatrix}
	\begin{pmatrix} m_{A'}^2 & m^2_{A' \varphi'} \\ m^2_{A' \varphi'} & m_{\varphi_n}^2 \end{pmatrix}
	\begin{pmatrix} \vec{A}' \\ \vec{\varphi}_n \end{pmatrix}
	\, = \,
	-\frac{1}{2}
	\begin{pmatrix} \vec{\tilde{A}}' & \vec{\tilde{\varphi}}_n \end{pmatrix}
	\begin{pmatrix} m_{\tilde{A}'}^2 & 0 \\ 0 & m_{\tilde{\varphi}_n}^2 \end{pmatrix}
	\begin{pmatrix} \vec{\tilde{A}}' \\ \vec{\tilde{\varphi}}_n \end{pmatrix},
\end{align}
where the off-diagonal element is defined as $m_{A' \varphi_n}^2 = \epsilon e \sqrt{2 m_b/(3 \pi)}\,R_{n00}(0)$. The mixing matrix diagonalizing the mass matrix and relating the two states $(\vec{A}', \vec{\varphi}_n)$ and $(\vec{\tilde{A}}', \vec{\tilde{\varphi}}_n)$ is
\begin{align}
	\begin{pmatrix} \vec{\tilde{A}}' \\ \vec{\tilde{\varphi}}_n \end{pmatrix} =
	\begin{pmatrix} \cos\vartheta_n & -\sin\vartheta_n \\ \sin\vartheta_n & \cos\vartheta_n \end{pmatrix}
	\begin{pmatrix} \vec{A}' \\ \vec{\varphi}_n \end{pmatrix}.
\end{align}
Here, the mass eigenstates and the mixing angle are obtained as $m_{\tilde{A}'}^2, m_{\tilde{\varphi}_n}^2 = \{(m_{A'}^2 + m_{\varphi_n}^2) \pm (m_{A'}^2 - m_{\varphi_n}^2)
[1 + 4m_{A' \varphi_n}^4/(m_{A'}^2 - m_{\varphi_n}^2)^2]^{1/2}\}/2$, and $\tan(2\vartheta_n) = 2 m_{A' \varphi_n}^2/(m_{\varphi_n}^2 - m_{A'}^2)$ with the domain of $-\pi/4 \leq \vartheta_n \leq \pi/4$. Then, the effective lagrangian in eq.\,(\ref{eq: pNR lagrangian with mixing modified bb 1}) is written as
\small
\begin{align}
    &\mathcal{L}^{(\text{pNR})}_V
    \simeq
    -\frac{1}{2} \vec{\tilde{A}}'\,(\Box + m_{\tilde{A}'}^2)\,\vec{\tilde{A}}'
    +\sum_i \bar{f}_i\,(i \Slash{\partial} + e\,Q_i\,\vec{A} \cdot \vec{\gamma} - m_{f_i})\,f_i
    -\frac{1}{2} \vec{\tilde{\varphi}}_n (\Box + m_{\tilde{\varphi}_n}^2) \vec{\tilde{\varphi}}_n
    \label{eq: pNR lagrangian with mixing modified bb 2}
    \\
    &-
    e \vec{\tilde{A}}' \cdot \sum_i Q_i \int d^4y
    \left[
        \epsilon \cos \vartheta_n\,\delta (x - y)
        - i e \sin \vartheta_n \sqrt{\frac{2m_b}{3\pi}}\,R_{n00}(0)\,{\cal G}^{(\gamma)}(x - y)
    \right] \bar{f}_i(y) \vec{\gamma} f_i(y)
    \nonumber \\
    &-
    e \vec{\tilde{\varphi}}_n \cdot \sum_i Q_i \int d^4y 
    \left[
        \epsilon \sin \vartheta_n\,\delta (x - y)
        + i e \cos \vartheta_n \sqrt{\frac{2m_b}{3\pi}}\,R_{n00}(0)\,{\cal G}^{(\gamma)}(x - y)
    \right] \bar{f}_i(y) \vec{\gamma} f_i(y)
    + \cdots,
    \nonumber
\end{align}
\normalsize
where "$\cdots$" involves other interactions, such as those of the bound states, $\vec{\varphi}_{m \neq n}(x)$.

\subsubsection{Decay width of the vector mediator particle}

To evaluate the total decay width of the vector mediator particle at the threshold region of $m_{A'} \sim 2 m_b$, we need knowledge of how the vector mediator particle and the upsilon mesons usually decay around this mass region. As seen in the lagrangian\,(\ref{eq: pNR lagrangian with mixing modified bb 1}), the mediator particle decays into a pair of SM leptons and that of SM quarks (or hadrons composed of the quarks) at leading order through the interaction $\epsilon e Q_i \vec{A}^\prime \cdot \bar{f}_i \vec{\gamma} f_i$. On the other hand, the upsilons have various decay channels at leading order: In addition to the decay channels into a pair of SM leptons and that of light SM quarks ($u$, $d$, $s$, $c$ quarks) through the effective interaction,
\small
\begin{align}
    e^2 Q_i \vec{\varphi}_n(x) \cdot \int d^4y\,{\cal G}^{(\gamma)}(x - y) \sqrt{\frac{2m_b}{3\pi}} R_{n00}(0)\,\bar{f}_i(y) \vec{\gamma} f_i(y)
    \sim
    \frac{e^2 Q_i}{\sqrt{3\pi}} \frac{R_{n00}(0)}{(2 m_b)^{3/2}}
    \vec{\varphi}_n \cdot \bar{f}_i \vec{\gamma} f_i,
    \nonumber
\end{align}
\normalsize
i.e., $\varphi_n \to f \bar{f}$, other decay channels into three gauge bosons, such as $\varphi_n \to g g g$ and $g g \gamma$, also contribute to the width significantly. Moreover, upsilons that are heavier than the lightest one decay into lighter upsilons through de-excitation transitions, $\varphi_n \to \varphi_m + X$\,($n > m$). Finally, the upsilon decays into a pair of B-mesons, i.e., $\varphi_n \to B \bar{B}$, when its mass is larger than $2m_B$, which dominates the decay width of the upsilon over the other channels.

We estimate the decay widths of the vector mediator particle in the following strategy. First, the partial decay width of the mediator particle into a pair of leptons at leading order, i.e., $\Gamma_0(A' \to \ell_i^- \ell_i^+)$, is already given in eq.\,(\ref{eq: perturbative decay into a lepton pair}). On the other hand, when $m_{A'} \simeq m_{\varphi_n}$, this decay width is modified due to the mixing effect. From the lagrangian\,(\ref{eq: pNR lagrangian with mixing modified bb 2}), it becomes
\begin{align}
    \tilde{\Gamma}_n(\tilde{A}' \to \ell^-_i \ell^+_i)
    \simeq \frac{\alpha m_{A'}}{3}
    \left[
        \epsilon \cos \vartheta_n + \frac{e}{\sqrt{3 \pi}}\frac{\sqrt{2m_b}\,R_{n00}(0)}{m_{A'}^2} \left( 1 - \frac{16}{3\pi} \alpha_s \right)^{1/2} \sin \vartheta_n
    \right]^2.
    \label{eq: decay width into leptons bb}
\end{align}
Here, we have involved the QCD correction in the above formula to reproduce the decay width\,(\ref{eq: branching fraction 1}), which is associated with the upsilon state (being now mixed with the mediator particle). The partial decay width into a pair of light ($u$, $d$, $c$, and $s$) quarks is also obtained in the same method, as mentioned in section\,\ref{subsec: perturbative decays}. The decay width at leading order is
\begin{align}
    \Gamma_0(A' \to q_i \bar{q}_i)
    \simeq \epsilon^2 \frac{\alpha m_{A'}}{3}
    \times 3\,Q_i^2\,(1 + \delta_{\rm QCD}),
    \label{eq: perturbative decay into a quark pair}
\end{align}
where $Q_i$ is the electric charge of the quark $q_i$. We have included the QCD correction in the above formula, which is associated with the $q_i \bar{q}_i$ final state and estimated to be $\delta_{\rm QCD} \simeq \alpha_s/\pi$ at one-loop level\,\cite{ParticleDataGroup:2020ssz}. As in the case of the decay into a lepton pair, the decay width in eq.\,(\ref{eq: perturbative decay into a quark pair}) is modified when $m_{A'} \simeq m_{\varphi_n}$. Its partial decay width is then obtained as follows:
\begin{align}
    \tilde{\Gamma}_n(\tilde{A}' \to q_i \bar{q}_i)
    =
    3\,Q_i^2\,(1 + \delta_{\rm QCD})\,\tilde{\Gamma}_n(\tilde{A}' \to \ell^-_i \ell^+_i).
    \label{eq: decay width into quarks bb}
\end{align}

In addition to the decay channels discussed above, when $m_{A'} \simeq m_{\varphi_n}$, the mediator particle also decays into three SM gauge bosons via the mixing with the upsilon mesons. Using the widths given in eqs.\,(\ref{eq: branching fraction 2}) and (\ref{eq: branching fraction 3}),  the partial decay widths of these processes are
\begin{align}
    \tilde{\Gamma}_n(\tilde{A}' \to \gamma g g)
    &
    \simeq \sin^2 \vartheta_n\,\Gamma[\Upsilon(nS) \to \gamma g g], \\
    \tilde{\Gamma}_n(\tilde{A}' \to g g g)
    & 
    \simeq \sin^2 \vartheta_n\,\Gamma[\Upsilon(nS) \to  3g],
\end{align}
with $\gamma$ and $g$ being a photon and a gluon, respectively.\footnote{The other decay channel into three gauge bosons is $\varphi_n \to \gamma \gamma \gamma$. We do not include this decay channel in our analysis, as its contribution to the decay widths of the upsilons is negligibly smaller than other channels.} Moreover, when $m_{A'} \simeq m_{\varphi_n}$ with $n \geq 2$, the mediator particle decays into a lighter $b \bar{b}$ bound state via a de-excitation process of the upsilon $\varphi_n$ by emitting light SM particles such as pion(s) and photon(s). Such processes significantly contribute to the total decay width of the mediator particle at $m_{A'} \simeq m_{\varphi_2}$ and at $m_{A'} \simeq m_{\varphi_3}$, so we involve these contributions to evaluate the decay width. Since no decay channels that significantly contribute to the widths of $\varphi_2$ and $\varphi_3$ exist other than those discussed so far, the sum of the partial decay widths of the de-excitation processes can be estimated using the total decay width of $\varphi_n$ presented in section\,\ref{subsec: decay into quarks} as follows:
\begin{align}
    \tilde{\Gamma}_n[\tilde{A}' \to (b\bar{b}) + X]
    \simeq \sin^2\vartheta_n
    \{\Gamma_n^{(bb)}
    &
    -\sum_i \Gamma\,[ \Upsilon(n S) \to \ell^-_i \ell^+_i ]
    -\sum_i \Gamma\,[ \Upsilon(n S) \to q_i \bar{q}_i ]
    \nonumber \\
    & \qquad\qquad
    -\Gamma[\Upsilon(nS) \to \gamma g g]
    -\Gamma[\Upsilon(nS) \to  3g] \},
\end{align}
with $n$ being two or three. Here, $(b\bar{b})$ represents a bound state composed of $b$ and anti-$b$ quarks, and $\Gamma_n^{(bb)}$ is the total decay width of the upsilon $\varphi_n$. Here, the partial decay width of $\Upsilon(nS)$ into a lepton pair, $\Gamma\,[ \Upsilon(n S) \to \ell^-_i \ell^+_i ]$, is the same as that in eq.\,(\ref{eq: branching fraction 1}), while the decay width into a quark pair $q_i\bar{q}_i$ is given by $\Gamma\,[ \Upsilon(n S) \to q_i \bar{q}_i ] = 3\,Q_i^2 (1 + \delta_{\rm QCD}) \Gamma\,[ \Upsilon(n S) \to \ell^-_i \ell^+_i ]$, as deduced from the above discussion. Finally, when the mass of the upsilon becomes larger than twice the $B$ meson mass, such as $\varphi_n$ with $n$ being more than three, i.e., the upsilons heavier than $\Upsilon(3S)$, it decays into a pair of the $B$ mesons at almost 100\% branching fraction. So, the sum of the partial decay widths of such processes can be approximated as
\begin{align}
    \tilde{\Gamma}_n[\tilde{A}' \to B \bar{B}]
    \simeq \sin^2 \vartheta_n\,\Gamma_n^{(bb)}. 
\end{align}

\begin{figure}[t]
    \centering
 \includegraphics[keepaspectratio, scale=0.35]{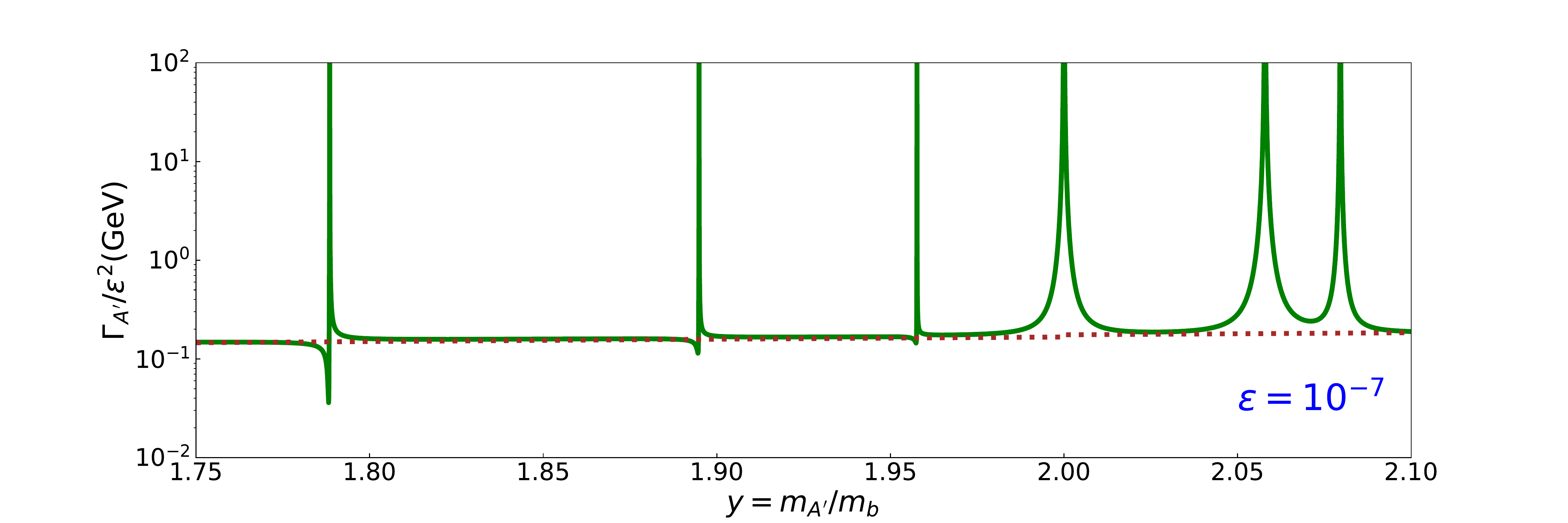}
    \caption{\small \sl A solid green line shows the total decay width of the mediator particle as a function of $y = m_{A'}/m_b$ with $m_b$ being 5.3\,GeV as given in section\,\ref{subsec: decay into quarks}. The $\epsilon$ parameter is fixed to be $10^{-7}$. The decay width obtained by the naive quark-parton model is also shown by a dotted brown line.}
    \label{fig: bottom mixing}
\end{figure}

With the above partial decay widths, the total decay width of the vector mediator particle in the vicinity of the threshold region of a bottom quark pair is obtained as follows:\footnote{The mediator particle also decays directly into final states composed of 3 gauge bosons, lighter $b \bar{b}$ bound states, and a pair of $B$ mesons via one-loop diagrams, etc. However, their widths are much smaller than those of the upsilons decaying into the same states. So, we do not include these contributions in the formula.}
\begin{align}
    \Gamma_{A'}
    &\simeq
    \sum_{\ell_i = e, \mu, \tau}
    \left\{
        \Gamma_0 (A' \to \ell_i^-\ell^+_i)
        + 
        \sum_n
        \left[
            \tilde{\Gamma}_n (\tilde{A}' \to \ell_i^- \ell_i^+)
            -
            \Gamma_0 (A' \to \ell_i^-\ell^+_i)
        \right]
    \right\}
    \label{eq: total decay width}
    \\
    &+
    \sum_{q_i = u, d, c, s}
    \left\{
        \Gamma_0 (A' \to q_i\bar{q}_i)
        + 
        \sum_n
        \left[
            \tilde{\Gamma}_n (\tilde{A}' \to q_i \bar{q}_i)
            -
            \Gamma_0 (A' \to q_i\bar{q}_i)
        \right]
    \right\}
    \nonumber \\
    &+
    \sum_n
    \left[
        \tilde{\Gamma}_n(\tilde{A}' \to \gamma g g)
        +
        \tilde{\Gamma}_n(\tilde{A}' \to g g g)
    \right]
    +
    \sum_{n= 2, 3}
    \tilde{\Gamma}_n[\tilde{A}' \to (b\bar{b}) + X]
    +
    \sum_{n \geq 4}
    \tilde{\Gamma}_n[\tilde{A}' \to B \bar{B}].
    \nonumber
\end{align}
The total decay width in eq.\,(\ref{eq: total decay width}) is depicted assuming $\epsilon = 10^{-7}$ in Fig.\,\ref{fig: bottom mixing} as a solid green line. The decay width obtained by the naive quark-parton model calculation is also shown by a dotted brown line. It is seen in the figure that the decay width is enhanced and suppressed\footnote{The suppression comes from the negative interference among the components in eqs.\,(\ref{eq: decay width into leptons bb}) and (\ref{eq: decay width into quarks bb}).} compared to that of the naive quark-parton model calculation in narrow $y$ regions where the mediator particle $A'$ has a sizable mixing with lighter upsilons $\varphi_n$ with $n \leq 3$. On the other hand, the decay width is enhanced in wider $y$ regions when $A'$ is primarily mixed with upsilons $\varphi_n$ with $n \geq 4$, as the heavier upsilons couple to the SM particles more strongly.

\subsubsection{Production of the vector mediator particle at \texorpdfstring{$e^- e^+$}{TEXT} colliders}

\begin{figure}[t]
    \centering
   \includegraphics[keepaspectratio, scale=0.35]{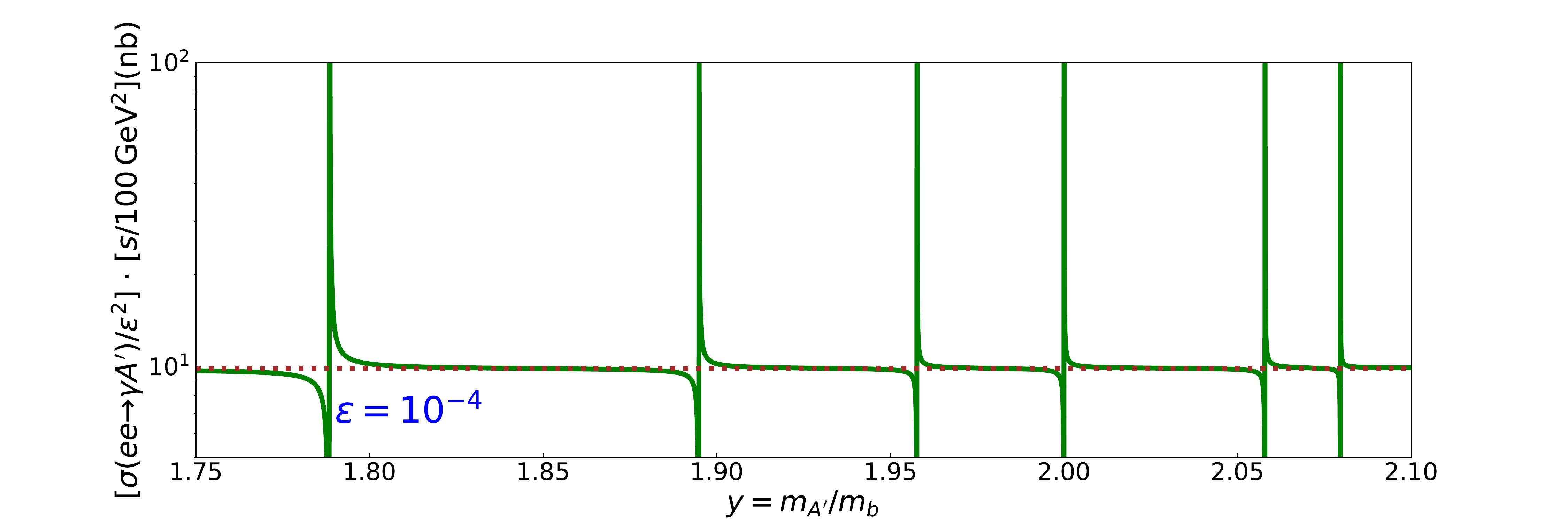}
    \caption{\small \sl The production cross-section of the mediator particle as a function of $y = m_{A'}/m_b$, obtained by integrating the differential cross-section\,(\ref{eq: production cross-section bb 2}) over the angles $-0.98 \leq \cos\theta \leq 0.98$. The cross-section obtained by the perturbative calculation is also shown by a dotted brown line.}
    \label{fig: production and events at bb}
\end{figure}

As in the case of the mixing with muonic atoms discussed in the previous subsection, the mixing with upsilons also affects the production of the mediator particle at collider experiments. At given center-of-mass energy $s^{1/2} \gg m_{A'}$, the mediator particle is efficiently produced at $e^- e^+$ colliders via the process, $e^- e^+ \to A' \gamma$. Its differential cross-section obtained by the perturbative calculation at leading order is given in eq.\,(\ref{eq: perturbative production cross-section mumu}). On the other hand, when the mediator particle degenerates with one of the upsilons, i.e., $m_{A'} \simeq m_{\varphi_n}$, the differential production cross-section is modified and obtained by the lagrangian\,(\ref{eq: pNR lagrangian with mixing modified bb 2}) as follows:
\begin{align}
    \frac{d \tilde{\sigma}_n (e^- e^+ \to \gamma \tilde{A}')}{d\cos\theta} \simeq
    \left[
        \epsilon \cos \vartheta_n
        +
        \frac{e}{\sqrt{3 \pi}}\frac{\sqrt{2m_b}\,R_{n00}(0)}{m_{A'}^2} \sin \vartheta_n
    \right]^2
    \frac{2 \pi \alpha^2}{s} \frac{1 + \cos^2\theta}{\sin^2\theta},
    \label{eq: production cross-section bb}
\end{align}
with $\theta$ being the angle between the directions of the mediator particle and the incoming electron. When the mixing angle $\vartheta_n$ is suppressed, the above production cross-section becomes the same as that of the perturbative calculation in eq.\,(\ref{eq: perturbative production cross-section mumu}). So, the production cross-section at the threshold region of a bottom quark pair (as a function of $m_{A'}$) is given by
\begin{align}
    \frac{d \sigma (e^- e^+ \to \gamma A')}{d\cos\theta}
    \simeq
    \frac{d \sigma_0 (e^- e^+ \to \gamma A')}{d\cos\theta}
    +
    \sum_n
    \left[
        \frac{d \tilde{\sigma}_n (e^- e^+ \to \gamma \tilde{A}')}{d\cos\theta}
        -
        \frac{d \sigma_0 (e^- e^+ \to \gamma \tilde{A}')}{d\cos\theta}
    \right].
    \label{eq: production cross-section bb 2} 
\end{align}
The production cross-section of the mediator particle, obtained by integrating the above differential cross-section over the scattering angles $-0.98 \leq \cos\theta \leq 0.98$, is depicted in Fig.\,\ref{fig: production and events at bb} assuming $\epsilon = 10^{-4}$. The cross-section is significantly affected when $m_{A'} \simeq m_{\varphi_n}$, as expected. The region that is influenced by the mixing is narrower when the mediator particle mixes with a bound state with a larger $n$, i.e., a heaver upsilon, as the corresponding value of the wave-function at origin, $R_{n00}(0)$, is smaller for a larger $n$, as seen in section\,\ref{subsec: decay into quarks}.\footnote{We have calculated the production cross-section using the narrow width approximation as seen in eq.\,(\ref{eq: production cross-section bb 2}), because the width of the mediator particle is still narrow even if it mixes significantly with upsilons. On the other hand, the effect of the finite width will slightly change the region influenced by the mixing in Fig.\,\ref{fig: production and events at bb}.} In addition to the production of the vector mediator particle discussed above (or in section\,\ref{subsubsec: production at mu mu}), that of the SM bound state degenerated with the mediator particle, i.e., the upsilon (or the muonic atom), is also affected if the mixing angle becomes large. So, the measurement of both the mediator particle and the bound state at the collider experiment is required to analyze the signal correctly. We will leave such a detailed analysis for future work.

\section{Summary and discussion}
\label{sec: summary}

We have developed the formula to quantitatively estimate the effect of the threshold singularity that emerged in the dark sector scenario with a light mediator. We take the minimal model of the vector mediator particle with a mass around twice the muon mass and twice the bottom quark mass as concrete examples. Using the potential non-relativistic lagrangian method, we derived the effective lagrangian, which describes the interaction between the vector mediator particle and the two-body state composed of a muon/bottom quark pair at the threshold region. The interaction between a muon pair is from the Coulomb force, so, in this case, we have derived the formula by explicitly solving the corresponding Schrodinger equation. On the other hand, in the case of a bottom quark pair, the interaction between the quarks is mainly from the strong force, and, unlike the QED case, the infrared QCD behavior complicates the description of the two-body state. So, we estimated the wave function describing the two-body state by utilizing the spectroscopic data of QCD bound states.

Using the effective lagrangian, we have calculated some physical quantities, such as the total decay width of the mediator particle, assuming that it decays visibly, i.e., mainly into SM particles. We have found that the decay width is slightly enhanced because of the Sommerfeld effect when the mass of the mediator particle is just above the threshold of a muon pair, i.e., twice the muon mass. On the other hand, when the mediator mass is in the vicinity of those of muonic atoms below the threshold, the decay width is strongly enhanced because of the mixing effect. Moreover, when the mass is around the threshold of a bottom quark pair, the width is more strongly enhanced, for the vector mediator particle mixes with upsilon mesons more that are strongly interacting with various SM particles.

Such an enhancement is expected to give interesting implications to collider physics and cosmology. First, concerning collider physics, enhancing the total decay width is equivalent to shortening the lifetime of the mediator particle. Furthermore, it also means strengthening the interaction with incident colliding particles. So, collider signals of the mediator particle are significantly altered compared to those without taking the singularity effect into account. At the threshold regions, known to be those experimentally uncharted yet, the formula developed in this paper must be mandatory to evaluate the collider signals correctly. On the other hand, concerning cosmology, enhancing the total decay width means increasing the strength of the interaction between the mediator particle and SM particles composing the thermal medium in the early universe. So, it affects the cosmology of the mediator particle differently from that without such an enhancement. Moreover, even if the mass of the mediator particle does not lie very close to the threshold, the thermal correction to the masses of the bound states renders the singularity effect active in the evolution of the universe, so the cosmology is influenced by the effect in a wide mass region of the mediator particle. We leave a detailed discussion of the implications for future studies.

\appendix

\section{Matching method for the decay width \texorpdfstring{$\Gamma\,(A' \to \mu^- \mu^+)$}{TEXT}}
\label{app: matching}

As discussed in section\,\ref{subsec: decay into leptons}, the partial decay width of the vector mediator particle into a muon pair at the threshold, $\Gamma\,(A'\to \mu^- \mu^+)$ in eq.\,(\ref{eq: decay width into leptons}), has the following form at $m_{A'} \geq 2m_\mu$:
\begin{align}
    \Gamma_1(A' \to \mu^- \mu^+)
    \equiv \frac{2 \pi \epsilon^2 \alpha^2 m_\mu^2 }{m_{A'}\,\left[1 - \exp\,(-\pi\alpha/\sqrt{m_{A'}/m_\mu - 2})\right]}.
\end{align}
At the vanishing limit of the long-range force, i.e., $\alpha \to 0$ (in the denominator), it gives
\begin{align}
    \Gamma_2(A' \to \mu^- \mu^+)
    \equiv \lim_{\alpha \to 0} \Gamma_1(A' \to \mu^- \mu^+)
    = \frac{2 \epsilon^2 \alpha m_\mu^2}{m_{A'}}\,\sqrt{m_{A'}/m_\mu - 2}.
\end{align}
Remember that the above partial decay width, obtained by the pNR lagrangian method, only involves the s-wave contribution to the width of the mediator particle. In contrast, the width obtained by a perturbative calculation in eq.\,(\ref{eq: perturbative decay into a lepton pair}) involves all the partial wave contributions to the width at leading order; we adopt the following method to match the decay widths obtained by the non-perturbative and the perturbative calculations:
\begin{align}
    \Gamma\,(A' \to \mu^- \mu^+)
    =
    \Gamma_1(A' \to \mu^- \mu^+) - \Gamma_2(A' \to \mu^- \mu^+) + \Gamma_0(A' \to \mu^- \mu^+).
    \label{eq: gamma mu mu at app}
\end{align}
We use this partial decay width to depict it in Fig.\,\ref{fig: decay width into leptons} at the mass region of $m_{A'} \geq 2m_\mu$. 

\section{Matching method for the decay width \texorpdfstring{$\Gamma\,(A' \to b \bar{b})$}{TEXT}}
\label{app: matching 2}

As discussed in section\,\ref{subsec: decay into quarks}, the "partial decay width" of the vector mediator particle into a pair of bottom quarks at the threshold region, $\Gamma\,(A'\to b \bar{b})$, has the following form:
\begin{align}
    \Gamma_{\rm Res}( A' \to b \bar{b} )
    = \sum_n \frac{2 \epsilon^2 \pi \alpha}{3 m_{A'}}|R_{n00}(0)|^2f(m_{A'};M^{(b b)}_n,\Gamma^{(b b)}_n/2).
    \label{eq: non-perturbative contribution}
\end{align}
It can be seen from the formula that only the bound states (discrete states) appear because of the effect of color confinement. As in the case of the decay into a muon pair, this width is obtained using the imaginary part of the self-energy of the mediator particle,
\begin{align}
    \Gamma(A' \to b\bar{b}) = -\frac{1}{m_{A'}}{\rm Im}\Pi(m_{A'}).
\end{align}
Actually, there are two different contributions to (the imaginary part of) the above self-energy. One is nothing but the one given in eq.\,(\ref{eq: non-perturbative contribution}), which is called the non-perturbative (resonance) contribution, $\Pi_{\rm Res}(q)$, induced by infinitely many exchanges of photons and gluons between the (non-relativistic) bottom quarks. Hence, it generates the resonances corresponding to the Upsilon mesons as discussed in section \,\ref{subsec: decay into quarks}. The other contribution is the continuum (non-resonant) ones, $\Pi_{\rm Per}(q)$, which gives a non-zero imaginary part of the self-energy that appears in the region of $q \geq 2\,m_B$ with $m_B $ being the mass of the B meson. This contribution can be computed within a perturbative QCD framework as follows:
\begin{align}
    {\rm Im}\,\Pi_{\rm Per}(q)
    &\simeq
    -\epsilon^2 \frac{\alpha }{9} q^2 \sqrt{1 - \frac{4m_b^2}{q^2}} \left(1 + \frac{2 m_b^2}{q^2}\right) [1 + \delta_{\rm QCD}(q)],
    \nonumber \\
    \delta_{\rm QCD}(q)
    &\simeq
    \sum^4_{n=1}c_n\left(\frac{\alpha_s(q)}{\pi}\right),
\end{align}
where $\delta_{\rm QCD}(q)$ is the QCD correction and is computed at 3-loop levels.\footnote{Here, we fix the value of $m_b$ to be 5.3\,GeV as that in eqs.\,(\ref{eq: branching fraction 2}), (\ref{eq: branching fraction 3}), which is almost the mass of the B meson. The detailed behavior of $\Pi_{\rm Per}(m_{A'})$ at $m_{A'} \simeq 2\,m_b$ is, however, hidden behind the $\Upsilon (4S)$ resonance.} See Ref.\,\cite{ParticleDataGroup:2020ssz} for the precise forms of the coefficients $c_n$s, as well as the QCD running coupling $\alpha_s(q)$. We, therefore, use the partial decay width of the mediator particle into a bottom quark pair,
\begin{align}
    \Gamma( A' \to b \bar{b} ) = \Gamma_{\rm Res}( A' \to b \bar{b} ) - \frac{1}{m_{A'}} {\rm Im}\,\Pi_{\rm Per}(m_{A'}),
\end{align}
instead of the width given in eq.\,(\ref{eq: decay width into hadrons}) itself, to depict the solid green line in Fig.\,\ref{fig: decay width into quarks}.

\bibliographystyle{unsrt}
\bibliography{refs}

\end{document}